\documentclass[aps,pra,onecolumn,superscriptaddress,longbibliography]{revtex4-2}
\usepackage{amsmath,amsfonts,amssymb}
\usepackage{bm,graphicx,braket}
\usepackage[usenames,dvipsnames]{color}
\usepackage[colorlinks,linkcolor=black,citecolor=blue,urlcolor=blue]{hyperref}

\renewcommand{\vec}[1]{\bm{\mathrm{#1}}}
\newcommand\Rv{\vec{R}}
\newcommand\rv{\vec{r}}

\newcommand{\dtau}{\delta_\tau}
\newcommand{\ih}{\widehat{i}}
\newcommand{\Rvh}{\widehat\Rv}
\newcommand{\rvh}{\widehat\rv}
\newcommand{\rvch}{\widehat{\rvc}}
\newcommand{\Xh}{\widehat{X}}
\newcommand{\compactcoord}{x}
\newcommand{\Compactcoord}{X}
\newcommand{\rc}{\compactcoord}
\newcommand{\rvc}{\vec{\compactcoord}}
\newcommand{\Rvc}{\vec{\Compactcoord}}
\newcommand{\Wv}{{\vec{W}}}
\newcommand{\wv}{{\vec{w}}}
\newcommand{\So}{{\mathcal{S}^1}}

\begin{document}

\author{G. Spada}
\affiliation{Dipartimento di Fisica, Universit\`a di Trento and CNR-INO BEC
 Center, 38123 Povo, Trento, Italy}
\author{S. Giorgini}
\affiliation{Dipartimento di Fisica, Universit\`a di Trento and CNR-INO BEC
 Center, 38123 Povo, Trento, Italy}
\author{S. Pilati}
\affiliation{School of Science and Technology, Physics Division, Universit\`a
 di Camerino, 62032 Camerino, Italy}
\affiliation{INFN-Sezione di Perugia, 06123 Perugia, Italy}

\title{Path-integral Monte Carlo worm algorithm for Bose systems \\ with
 periodic boundary conditions}
\begin{abstract}
 We provide a detailed description of the path-integral Monte Carlo worm
 algorithm used to exactly calculate the thermodynamics of Bose systems in the canonical
 ensemble. The algorithm is fully consistent with periodic boundary conditions,
 that are applied to simulate homogeneous phases of bulk systems, and it does not
 require any limitation in the length of the Monte Carlo moves realizing the sampling
 of the probability distribution function in the space of path configurations.
 The result is achieved adopting a representation of the path coordinates where
 only the initial point of each path is inside the simulation box, the remaining
 ones being free to span the entire space. Detailed balance can thereby be
 ensured for any update of the path configurations without the ambiguity of
 the selection of the periodic image of the particles involved.
 We benchmark the algorithm using the non-interacting Bose gas model for which
 exact results for the partition function at finite number of particles can be derived.
 Convergence issues and the approach to the thermodynamic limit are also addressed
 for interacting systems of hard spheres in the regime of high density.
\end{abstract}
\maketitle

\section{Introduction}

The path-integral Monte Carlo (PIMC) method is a computational approach
that allows one to exactly calculate the equilibrium properties of Bose systems at
finite temperature starting from the microscopic Hamiltonian. The first
applications of the method addressed the study of bulk liquid and solid $^4$He
approaching the quantum degenerate regime~\cite{pollock1984simulation,
 PhysRevLett.56.351}. Similarly to these first simulations, many later
implementations of the PIMC algorithm addressed homogeneous systems featuring
infinite spatial extension in the physically relevant dimensions. To mimic such
configurations, the use of periodic boundary conditions in the computer
simulations is crucial~\cite{ceperley1995path}. The size of the simulation cell
enters as an important parameter of the numerical procedure, calling for a
careful extrapolation of all the results to a properly defined thermodynamic
limit.
Another crucial aspect of PIMC simulations is the efficient sampling of
Bose-particle permutations. This gets increasingly important with the level of
quantum degeneracy and is essential to obtain reliable estimates of observables
such as the superfluid density and the condensate
fraction~\cite{pollock1987path,Boninsegni2005}. In this context, an important
technical advancement emerged with the introduction of the \emph{worm
 algorithm}, first devised for lattice models~\cite{prokof1998worm,
 prokof2001worm} and later extended to continuous-space
simulations~\cite{boninsegni2006worma, boninsegni2006wormb}. The PIMC method
implementing the worm algorithm has proven to be one of the most powerful
computational quantum many-body techniques. It allowed performing accurate
simulations of intriguing phenomena in different condensed matter systems, such
as dipolar systems~\cite{PhysRevLett.119.215302, Cinti2017,
 PhysRevA.100.063614}, ultracold gases~\cite{PhysRevLett.105.135301,
 PhysRevB.83.092506, PhysRevLett.111.050406, Cinti2014DefectinducedSW,
 PhysRevLett.127.205301} and quantum fluids and
solids~\cite{PhysRevLett.96.105301, PhysRevLett.97.080401, PhysRevB.78.245414,
 PhysRevB.79.174203, Rota2011PathIM, PhysRevB.85.224513, PhysRevB.93.104510}.
However, the original implementation of the worm algorithm is not fully
compatible with periodic boundary conditions. This can lead to biased results
when the de Broglie thermal wavelength starts to be comparable to the
size of the fundamental periodic cell.
The basic idea behind the worm algorithm is the use of both diagonal and
off-diagonal configurations of the paths describing particles in the many-body
system. Various moves of portions of the paths are devised to ensure the
ergodic sampling of both types of configurations as well as to switch from
the diagonal to the off-diagonal sector and vice-versa. Such moves, satisfying
detailed balance, have a straightforward implementation in the case of a system
with infinite extension or of a confined finite geometry. In periodic systems,
instead, the ambiguity in the choice of periodic images might lead to
biased results, unless stringent constraints are imposed on the portions of
the paths involved in certain Monte Carlo moves. Such bias
can be avoided by adopting periodic cells much larger than the thermal
wavelength, but this becomes impractical in the zero temperature limit.
Furthermore, the limitations in the updates can affect the efficiency of
the Monte Carlo sampling.
We present here a formalism and a detailed recipe for a PIMC algorithm for
bosons in the canonical ensemble which is rigorously compatible with periodic
boundary conditions. With this algorithm, no constraint has to be imposed on
the Monte Carlo updates, even for small periodic cells. We use as a benchmark
the canonical non-interacting Bose gas in a periodic box for which exact
results for the energy can be obtained for any number of particles in the box.
Direct comparison with these results permits us to check one by one the various
elementary moves of our novel implementation of the worm algorithm, verifying
that they provide an unbiased ergodic sampling.
Interacting systems are considered only with the hard-sphere model interaction
and the pair-product approximation. In this case, no exact result is available
for finite box-like geometries even for just a pair of particles. Nonetheless,
we can investigate the convergence of the results for a given number of
particles in terms of the length of the path steps in imaginary time as well as
the approach to the thermodynamic limit when the number of particles in the
simulation is increased.\\

The structure of the paper is as follows. In section~\ref{sec:II} we introduce
the representation of the particle coordinates consistent with the use of periodic
boundary conditions. Trying to be pedagogical we first consider the case of a
single particle in a one-dimensional periodic box and then move to the case of
$N$ identical particles. In the same section we describe the general scheme of the
worm algorithm and we introduce in details the various Monte Carlo updates. In
section~\ref{sec:III} we benchmark the algorithm against exact results of the
non-interacting Bose gas for one, two and many particles at different
temperatures. Section~\ref{sec:IV} is devoted to interacting systems for which
we use the hard-sphere model in the regime of high density. We address the
convergence of the algorithm for a given number of particles $N$ and for
different temperatures as well as the approach to the thermodynamic limit.
Finally in the last section we draw our conclusions. In the appendix we
outline the useful formulas used for the calculation of the internal energy,
both the thermodynamic and virial estimator, and of the pressure.

\section{Path Integral Monte Carlo}
\label{sec:II}

In a PIMC simulation one aims at calculating the partition function $Z_N$ of a
Bose system of $N$ identical particles described by the Hamiltonian $H$ and
with inverse temperature $\beta=1/(k_BT)$, where $k_B$ is Boltzmann's constant.
In the coordinate representation, the partition function is defined as the trace
over all states $|\Rv\rangle$ of the density matrix
$\rho(\Rv,\Rv^\prime,\beta)=\langle\Rv|e^{-\beta H}|\Rv^\prime\rangle$
properly symmetrized,
\begin{equation}
 Z_N=\frac{1}{N!}\sum_P \int d\Rv\, \rho(\Rv,P\Rv,\beta) \;.
\end{equation}
Here, $\Rv=(\rv_1,\rv_2,...,\rv_N)$ collectively denotes the position
vectors of the particles and $P\Rv=(\rv_{p(1)},\rv_{p(2)},...,\rv_{p(N)})$
corresponds to the position vectors with permuted labels.
Furthermore, the sum in the above equation extends over the
$N!$ permutations of the particle labels. The partition function can be
conveniently rewritten in terms of the convolution integral
\begin{eqnarray}
 Z_N =\frac{1}{N!}\sum_P \int d\Rv \int d\Rv_1 \dots \int d\Rv_{M-1}
 \,\rho(\Rv,\Rv_1,\dtau) \dots
 \rho(\Rv_{M-1},P\Rv,\dtau) \;,
\end{eqnarray}
where $\dtau=\beta/M$. Starting from the above decomposition, the
calculation is mapped to a classical-like simulation of polymeric chains with a
number of beads $M$ equal to the number of terms of the convolution integral.
More specifically, one makes use of suitable approximations for the density
matrix	$\rho(\Rv,\Rv^\prime,\dtau)$ at the higher temperature
$1/\dtau$ and performs the  multidimensional integration over $\Rv,
 \Rv_1,\dots,\Rv_{M-1}$ as well as the sum over permutations $P$ by Monte Carlo
sampling~\cite{pollock1984simulation, PhysRevLett.56.351}. The whole procedure
can be unambiguously followed if the particle coordinates $\rv_1,\dots,\rv_N$
can span the entire space, for example when the Hamiltonian includes a
confining external potential. In the case of interest of a bulk system, where
the simulation cell corresponds to a box periodically replicated in space, the
coordinates $\rv_i$ can either denote the position of the $i$-th particle in
the box or of one of its periodic images in the neighboring boxes. This
ambiguity has troublesome consequences in the proper sampling of configurations
$\Rv$ with the correct probability distribution. In particular, a na\"ive
implementation of the Monte Carlo updates might lead to the violation of the
detailed balance condition.
In the following subsection we motivate the use of a specific
representation for
the particle coordinates that allows one to unambiguously construct a PIMC
algorithm that is consistent with periodic boundary conditions. The upshot is
that one should consider coordinates as belonging to the infinite
space throughout the whole simulation, and
invoke periodic
boundary conditions only when considering the periodicity of the trajectories
in imaginary time or the interaction among different particles. First, we
consider the simplest possible example, namely a single particle in one
dimension, to show how this coordinate representation naturally emerges in the
context of the path-integral representation of the partition function. Once the
single particle case has been clarified, we extend the formalism to the
$N$-body system with periodic boundary conditions.

\subsection{Path Integral for one particle with periodic boundary conditions}

We consider a particle moving on a circle $\mathcal{S}^1 \sim \mathbb{R}
 ~/~\mathbb{Z}$ of length $L$ whose momentum is quantized in units of $2\pi
 \hbar/L$. A position eigenstate on the circle can be represented as the
Fourier series
\begin{equation}
 \ket{\rc}_\So
 = \frac{1}{\sqrt{L}} \sum_{n=-\infty}^{\infty} e^{-\dfrac{i p_n \rc}{\hbar}}
 \ket{p_n}\,,
 \qquad
 p_n = \frac{2\pi \hbar}{L}n \,,
\end{equation}
but it can also be expressed as the superposition of eigenstates in the  space
$\mathbb{R}$ as
\begin{equation}
 \ket{\rc}_\So
 = \sqrt{\frac{L}{2\pi\hbar}} \sum_{n=-\infty}^{\infty} \ket{\rc + n L} \,.
\end{equation}
In the above equations the coordinate $\rc$ is limited to the fundamental cell
$[0,L)$ and we used the notation $\ket{}_\So$ to represent a state on $\So$
while with the standard ket $\ket{}$, without subscripts, we represented a
state on the real line $\mathbb{R}$. For later convenience we introduce the
following alternative labeling of the states on $\mathbb{R}$:
\begin{equation}
 \ket{\rc + n L} \equiv \ket{\rc,n}\,,
\end{equation}
which explicitly separates the coordinate in the fundamental cell from the
integer identifying one of the periodic images.
The momentum eigenstate $\ket{p_n}$ does not have this distinction since the
plane-wave decomposition can equivalently be expressed in the two spaces as
\begin{equation}
 \ket{p_n} = \frac{1}{\sqrt{L}} \int_0^L \hspace*{-5pt }d\rc \,
 e^{\dfrac{i p_n \rc}{\hbar}} \ket{\rc}_\So
 = \frac{1}{\sqrt{2\pi \hbar}} \int_{-\infty}^\infty \hspace*{-5pt }dx \,
 e^{\dfrac{i p_n x}{\hbar}} \ket{x} \,.
\end{equation}
We can then write the partition function of a free particle of mass $m$
in one
dimension with periodic boundary conditions using the set of coordinates on the
real line $\mathbb{R}$:
\begin{equation}
 Z_1 = \int_0^L \hspace*{-5pt }d\rc \, \sum_{n=-\infty}^{\infty} \bra{\rc,0}
 e^{- \dfrac{\beta p^2}{2m}} \ket{\rc, n}\,,
 \label{eq:Z1}
\end{equation}
where $p$ is the momentum operator. Notice that the above equation
computes the trace over the
Boltzmann factors by fixing a reference in the fundamental cell and then
summing over all periodic images. The reader can easily verify that the above
expression evaluates to the same result as the more familiar formula
$Z_1=\sum_n \exp[-\beta p_n^2/(2m)]$. It is also important to notice that
if one limits to $n=0$ the summation over images in eq.~\eqref{eq:Z1},
i.e.~only the particle in the simulation box is considered, the
result for $Z_1$ would be correct only in the limit $\lambda_T\ll L$, where
$\lambda_T=\sqrt{2\pi\hbar^2\beta/m}$ is the thermal wavelength. For the
particular example of a single particle in a box, this issue explains the
difficulty of a path-integral algorithm with periodic boundary
conditions~\cite{ceperley1995path}.
We can then define $\dtau = \beta / M$, with $M$ being a positive integer, and
introduce $M-1$ completeness relations on the real line  $\mathbb{R}$,
\begin{equation}
 1 = \int_{-\infty}^\infty dx \, \ket{x}\bra{x}\,,
\end{equation}
to obtain a representation of $Z_1$ suitable for a PIMC simulation:
\begin{equation}
 Z_1 = \int_0^L \hspace*{-5pt }d\rc \, \sum_{n=-\infty}^{\infty}
 \int_{-\infty}^\infty dx_1 \dots
 dx_{M-1} \,
 \bra{\rc,0} e^{- \frac{\dtau p^2}{2m}} \ket{x_1}
 \bra{x_1} \dots \ket{x_{M-1}}
 \bra{x_{M-1}} e^{- \frac{\dtau p^2}{2m}}\ket{\rc, n}\,.
\end{equation}
Notice that the expression above involves a product of matrix elements
obtained from states defined on the real line~$\mathbb{R}$. The periodicity of
the space just constrains the leftmost bra to span the fundamental cell, and
the rightmost ket to coincide with one of the images labeled by $n$.
Before moving to the more general case of a system of $N$ particles in
$D$-dimensions, let us introduce the analog of the partition function in
eq.~\eqref{eq:Z1} for non-diagonal (or two-point) configurations,
\begin{equation}
 G_1 = \frac{1}{L}\int_0^L \hspace*{-5pt }d\rc' \int_0^L \hspace*{-5pt }d\rc \,
 \sum_{n=-\infty}^{\infty} \bra{\rc,0} e^{- \dfrac{\beta p^2}{2m}} \ket{\rc',
  n}
 \,,
\end{equation}
where the matrix element is between a state in the fundamental cell and a
generic other state. This quantity will characterize the simulation in the
off-diagonal sector and it can be rewritten as
\begin{equation}
 G_1 =
 \frac{1}{L} \int_{-\infty}^{\infty} \hspace*{-5pt }dx' \int_0^L \hspace*{-5pt
 }d\rc \, \bra{\rc,0} e^{- \dfrac{\beta p^2}{2m}} \ket{x'}\,.
 \label{eq:G1}
\end{equation}
The generalizations of eqs.~\eqref{eq:Z1} and \eqref{eq:G1} to the
$D$-dimensional $N$-particle system, in which the paths of the particles
in imaginary time are represented
on the infinite space, with only one reference coordinate in the
fundamental cell,
provides an unambiguous parametrization of the PIMC method and will be
employed below to
construct a worm algorithm that is consistent with periodic boundary
conditions.

\subsection{Path Integral for $N$ bosons with periodic boundary conditions}

We consider now a system composed of $N$ identical bosons of mass $m$ contained
in a $D$-dimensional hypercube of volume $V = L^D$. The use of periodic
boundary conditions makes the space of coordinates the $D$-dimensional torus $(
 \mathbb{R} ~/~ \mathbb{Z} )^D$. Generalizing the notation of the previous
subsection, we introduce the coordinates inside the box $\Rvc = (\rvc_1,
 \rvc_2, \dots, \rvc_N)$, with $\rvc_i \in [0,L)^D$, and a
collection of integers
$\Wv
 = (\wv_1, \wv_2, \dots, \wv_N)$, with $\wv_i \in \mathbb{Z}^D$ labeling the
image of the $i$-th particle. An $N$-particle state can then be defined as
\begin{equation}
 \ket{\Rv = \Rvc + \Wv L} \equiv \ket{\Rvc, \Wv} \,.
\end{equation}
The partition function can then be written as
\begin{equation}
 Z_N =	\frac{1}{N!} \sum_P\sum_\Wv
 \int_{V} \; d\Rvc \,
 \bra{\Rvc, \vec{0}} e^{-\beta H} \ket{P\Rvc, \Wv} \,,
 \label{eq:Z_def}
\end{equation}
where $P\Rvc= (\rvc_{p(1)}, \rvc_{p(2)}, \dots, \rvc_{p(N)})$ indicates the
position vector in the box $V$ with permuted indices. By using the convolution
rule and introducing $M-1$ intermediate beads each at the inverse
temperature $\dtau = \beta/M$, the partition function can be written as
\begin{equation}
 Z_N
 = \frac{1}{N!} \sum_P
 \int \prod_{j=0}^{M-1} d\Rv_j \,
 \rho(\Rv_j,\Rv_{j+1},\dtau) \,,
 \qquad
 \Rv_M \equiv P \Rvc_0 + \Wv L\,.
 \label{eq:Z_def2}
\end{equation}
In the above equation we have combined the integration over $\Rvc_0$ and the
sum over the integers $\Wv$ into an integration over $\Rv_0$, and implicitly
required that the vector $\Rv_0$ appearing as the argument of the density
matrix has to be considered as the image in the fundamental cell,
i.e.~the vector $\Rvc_0$. Similarly to the one-dimensional case in the previous
subsection, the remaining
vectors $\Rv_1,\dots,\Rv_M$, span the entire space.
As previously stated, in a PIMC calculation one performs a classical-like
simulation of polymers formed by $M$ links, connecting $M+1$ \emph{beads}. Each
polymer corresponds to one boson, that can  close on itself (if $p(i)=i$) or on
another particle specified by the permutation index $p(i)$. Notice that each
polymer $i$ starts at the bead $0$ inside the fundamental cell, and ends at
bead $M$, closing on an image of the first bead of the polymer $p(i)$,
identified up to the periodicity of the $D$-dimensional torus.
Denoting with $X$ a configuration of the $N$ particles (that will be
defined more in detail later), one can see that eq.~\eqref{eq:Z_def2}
naturally provides a probability density function (PDF), given by
\begin{equation}
 \pi(X) = \frac{1}{Z_N} \frac{1}{N!}
 \prod_{j=0}^{M-1}
 \rho(\Rv_j,\Rv_{j+1},\dtau) \,,
 \label{eq:pi_Z}
\end{equation}
that can be sampled using the Metropolis--Hastings
algorithm~\cite{metropolis1953equation,hastings}. Diagonal thermodynamic
observables can then be directly computed using the above PDF from the
configurations visited by the simulation. With obvious notation, the
equilibrium statistical average of a given operator $O$ is given by
\begin{equation}
 \braket{O} = \int \mathcal D X \, \pi(X) O(X) \,.
 \label{eq:Op}
\end{equation}

For the systems that are considered here, the Hamiltonian can be expressed as
the sum $H=\mathcal{T} + \mathcal{V}$ of a quadratic kinetic part
$\mathcal{T}=\sum_i p_i^2/(2m)$, with $p_i$ being the momentum operator
associated to particle $i$, and a potential part $\mathcal{V}$, usually
diagonal in the coordinate representation. In such a case, for sufficiently
large $M$ an accurate approximation of the high-temperature density matrices is
ensured by the Trotter product formula~\cite{trotter1959product}
\begin{equation}
 e^{-\beta(\mathcal{T} + \mathcal{V})}
 = \lim_{M\to\infty} \left[
 e^{-\dtau \mathcal{T}}e^{-\dtau \mathcal{V}}\right]^M \,,
\end{equation}
allowing us to rewrite the density matrices appearing in eq.~\eqref{eq:pi_Z} as
\begin{equation}
 \rho(\Rv_j, \Rv_{j+1}, \dtau)= \rho_\mathrm{free}(\Rv_j, \Rv_{j+1}, \dtau)
 \exp{\left[ - U(\Rv_j, \Rv_{j+1})\right]}\,,
 \label{eq:rho_fact}
\end{equation}
where $U$ is the potential energy term, while $\rho_\mathrm{free}$ is the free
particle density matrix obtained from the kinetic operator
\begin{equation}
 \rho_\mathrm{free}(\Rv_j, \Rv_{j+1}, \dtau)
 \equiv \prod_{i=1}^{N} \rho_\mathrm{free}^\mathrm{sp}(\rv_{i,j}, \rv_{i,j+1},
 \dtau)
 = \prod_{i=1}^{N} (4\pi\lambda\dtau)^{-D/2}\exp\left[-\frac{ (\rv_{i,j}
   -\rv_{i,j+1})^2}{4\lambda\dtau}\right] \,,
 \label{eq:rho_free}
\end{equation}
with $\lambda = \hbar^2/(2m)$. In the so-called \emph{symmetrized primitive
 approximation} the potential energy term is given by $U(\Rv_j, \Rv_{j+1}) =
 \dtau \left(\mathcal{V}(\Rv_{j})  + \mathcal{V}(\Rv_{j+1}) \right)/2$.
Depending on the problem at hand, one might significantly reduce the number $M$
of slices needed for convergence using different approximation schemes. An
effective scheme in the case of dilute systems with hard-sphere interaction is
the pair-product ansatz \cite{ceperley1995path}. This is discussed in more
detail in section~\ref{sec:IV}.
Let us anticipate that the factorization of eq.~\eqref{eq:rho_fact} highlights
the possibility of using efficient strategies to exactly sample the free
Gaussian part $\rho_\mathrm{free}$, leaving the acceptance/rejection stage of
the Metropolis--Hastings algorithm to be determined only by the potential
interaction term. Among the possible schemes we adopt the staging algorithm
\cite{sprik1985staging,sakkos}, a smart collective displacement of an arbitrary
number of beads.

\subsection{Worm algorithm}

The configuration space detailed above, which will be called $Z$-sector in what
follows, consists of polymers organized by the permutation vector $P$ into a
number of cycles, i.e.~subsets with cyclic permutations, meaning that each
polymer has links connecting it to a preceding and a following polymer. Instead
of directly summing the $N!$ integrals corresponding to just as many
permutations, we use a Monte Carlo integration strategy based on the worm
algorithm~\cite{prokof1998exact,prokof1998worm,prokof2001worm,boninsegni2006worma,boninsegni2006wormb},
which is based on an extended configuration space where the $Z$-sector is
augmented by a $G$-sector, obtained by cutting one of the cycles and leaving
one sequence open-ended. This open sequence of polymers constitutes the
\emph{worm}, with the first polymer called the \emph{tail} and the last one
called the \emph{head}. During the simulation, the system randomly fluctuates
between the two sectors by opening or closing the worm and, when in the
$G$-sector, the head of the worm might be swapped with another polymer,
thus sampling the permutations and allowing the creation of long permutation
cycles.
Configurations in the $G$-sector are obtained from a configuration in the
$Z$-sector complemented by a particle index $i_H$, indicating the head of the
worm, and an extra position vector, $\rv_{i_H,M} = \rvc_H + \wv_{i_H} L$,
representing an additional bead at the head of the worm dangling at the time
slice $j=M$. We need this additional position vector because we want the PDF~in
the $G$-sector to be the analogue of the one in the $Z$-sector, thus requiring
the same total number of links.
We can now define the analogue of the partition function in the
$G$-sector as
\begin{equation}
 G_N
 = \frac{1}{V N!} \sum_P \sum_{i_H=1}^N
 \int_V \; d\rvc_H\int \prod_{j=0}^{M-1} d\Rv_j \,
 \rho(\Rv_j,\Rv_{j+1},\dtau) \,,
 \quad
 \text{with~~}
 \rv_{i,M} =
 \begin{cases}
  \rvc_{p(i),0} + \wv_i L & \text{~~for~~} i \neq i_H \,, \\
  \rvc_H + \wv_{i_H} L    & \text{~~for~~} i = i_H \,,    \\
 \end{cases}
 \label{eq:ZG_def}
\end{equation}
where the $1/V$ factor has been introduced for dimensional reasons and, as
before, we have hidden the summation over the integers $\Wv$ inside the
integration over $\Rv_0$. Notice that, with respect to the partition function
in eq.~\eqref{eq:Z_def2}, the above equation contains the additional
integration over the vector $\rvc_{H}$ and the sum over the head index $i_H$,
representing the $N$ possible choices for cutting the permutation cycles.
We can then combine $Z_N$ and $G_N$ into a generalized partition function
\begin{equation}
 Z^\mathrm{worm}_N = Z_N + C G_N \,,
 \label{eq:ZW_def}
\end{equation}
where $C$ is an arbitrary constant that controls the relative simulation time
spent in the two sectors.
Denoting with $N_G$ and $N_Z$ the number of times the simulation is found in
the $G$ and $Z$-sector respectively, the
proportionality
\begin{equation}
 \frac{N_G}{N_Z} = C \frac{G_N}{Z_N} \,,
 \label{eq:NG_NZ}
\end{equation}
must be satisfied, implying that the parameter $C$ can be tuned to
optimize the simulation.
Indeed, the statistical autocorrelation of observables is minimized if
$N_G \sim N_Z$.
Using eqs.~\eqref{eq:Z_def2} and \eqref{eq:ZG_def} we can rewrite
eq.~\eqref{eq:ZW_def} as
\begin{equation}
 \begin{split}
  Z^\mathrm{worm}_N =  \sum_{S =Z,G}
  \sum_P \sum_{i_H=1}^N
  &\int d\rvc_H \prod_{j=0}^{M-1} d\Rv_j
  \times \\
  \times
  \frac{1}{V N!}
  &\left(
  \prod_{j=0}^{M-2} \rho(\Rv_j,\Rv_{j+1},\dtau) \right)
  \left\lbrace
  \frac{\delta_{S,Z}}{N}\,
  \rho(\Rv_{M-1},\Rv_M,\dtau)
  + C \delta_{S,G}\,
  \rho(\Rv_{M-1},\Rv_M^H,\dtau)
  \right\rbrace\,,
 \end{split}
 \label{eq:ZW}
\end{equation}
where we have introduced an explicit summation over the two sectors as well as
the Kronecker delta functions to select the proper integrand for each sector.
With the notation $\Rv_M^H$ we represent the vector with components $\rv_{i,M}=
 \rvc_{p(i),0} + \wv_i L$ for $i \neq i_H$ and with component $\rv_{i_H,M} =
 \rvc_H + \wv_{i_H} L$ for $i = i_H$. A generic configuration $X$ for the worm
algorithm is specified as $X = (S,P,i_H,\rvc_H,\lbrace \Rv_j \rbrace)$ with $S$
the sector, $P$ the permutation vector, $i_H$ the head index, $\rvc_H$ the
extra head vector and $\lbrace \Rv_j \rbrace$ the set of $N M$ position vectors
of the polymers.
The PDF~for the worm algorithm is readily obtained from eq.~\eqref{eq:ZW} and
reads
\begin{equation}
 \pi_W(X) =
 \frac{1}{Z^\mathrm{worm}_N}
 \frac{1}{V N!}
 \left(
 \prod_{j=0}^{M-2} \rho(\Rv_j,\Rv_{j+1},\dtau) \right)
 \left\lbrace
 \frac{\delta_{S,Z}}{N} \,
 \rho(\Rv_{M-1},\Rv_{M},\dtau)
 + C \delta_{S,G} \,
 \rho(\Rv_{M-1},\Rv_{M}^H,\dtau)
 \right\rbrace\,.
\end{equation}
For sufficiently large number of beads $M$ the density matrices are then
further factorized into a free particle term and an interaction term as in
eq.~\eqref{eq:rho_fact}.
Before turning to the implementation of the Monte Carlo algorithm we point out
that the worm algorithm offers one additional advantage besides the efficient
sampling of permutations: it allows accessing the $G$-sector configurations
(also called off-diagonal configurations) from which one can extract other
useful observables such as a properly normalized one-body density
matrix~\cite{boninsegni2006wormb}.

\subsection{Monte Carlo updates}

The Monte Carlo procedure is based on a random walk and is obtained by means of
semi-local updates $X \to X'$ in the configuration space. The transition matrix
$P(X, X')$, i.e.~the probability to go from the state $X$ to $X'$, is chosen
such that it satisfies the detailed balance condition
$\pi_W(X) P(X, X') = \pi_W(X') P(X', X)$.
Together with the ergodicity condition, this ensures that,
after equilibration, the random walk samples points with the probability
$\pi_W(X)$.
According to the  Metropolis--Hastings criterion, the transition probability for
$X \neq X'$ can be factorized into an a priori sampling distribution $T(X, X')$
and an acceptance probability $A(X, X')$ as $P(X, X') = T(X, X') A(X, X')$. The
trial moves are then accepted (or rejected) according to the probability:
\begin{equation}
 A(X, X') = \min \left[ 1, \frac{T(X', X)\pi_W(X')}{T(X, X')\pi_W(X)}
  \right]\,.
 \label{eq:A_metropolis}
\end{equation}
We provide below the details for an efficient set of moves (translate, redraw,
open/close, move head, move tail) that allow the sampling of the probability
distribution $\pi_W$.
Before we start, let us stress once again that in order to develop a code
compatible with the periodic boundary conditions one must consistently use the
coordinates as defined on $\mathbb{R}^D$ throughout the
simulation. The only instance in which one needs to use the
representation of the periodic images is to ``recenter'' a polymer when the
initial bead is moved outside of the fundamental cell: in that case one rigidly
translates the whole polymer by the factor $\Delta\wv L$, with $\Delta\wv$
chosen to make the polymer start in the fundamental cell.
We note that in the simulation one can either choose to use the integers $\Wv$
or to use the additional vector $\Rv_M$, making sure that $\Rv_M = P \Rvc_0 +
 \Wv L$ in the $Z$-sector and $\Rv_M = \Rv_M^H$ in the $G$-sector. We find
the second choice more convenient because it provides a uniform
representation of the polymers, which better suits a computer program.
A separate discussion is needed in the case of the computation for the
interaction term. For systems with pairwise interactions, one needs to compute
the distance, in the compactified space, between beads belonging to different
polymers. Different approaches might be used in this case, but for interactions
that decay sufficiently fast with the distance, one can effectively use the
\emph{nearest image} convention, in which the distance along each direction is
computed modulo $L$.
In the computation of the energy terms for a given link, say, from time
slice $j$ to $j+1$, we identify the closest periodic image of a bead at $j$,
and then find the corresponding image of the subsequent bead at $j+1$.\\

Before describing the various Monte Carlo updates, we should emphasize
that standard implementations of the worm algorithm had to put
constraints on the length of the path segments, namely the number of beads,
involved in the open/close and swap moves, which had to be much smaller than
the size $L$ of the box to avoid biased sampling. Apart from adding an
inconvenient issue related to the fine tuning of the parameters of the
simulation, such constraints reduce in general the efficiency of the Monte
Carlo sampling.

\subsubsection{Translate}

This update translates all polymers belonging to a permutation cycle as a rigid
body. We select a particle index from a uniform random distribution and
construct a list of all the particles in the same permutation cycle. We then
select a displacement vector $\Delta \rv$ by sampling $D$ random variables
uniformly between $0$ and a maximum displacement $r_{\mathrm{max}}$ and we
perform the shift $\rv^\prime_{i,j} = \rv_{i,j} + \Delta \rv$ for all the beads
$j=0,\dots,M$ and all the particles $i$ belonging to the permutation cycle.
This shift keeps the internal links of all polymers fixed, hence the
probability to accept this update is given by
\begin{equation}
 A_T = \min \left\lbrace 1,
 \exp \left[
  \sum_{j=0}^{M-1} \left(  U(\Rv_j,\Rv_{j+1}) -
  U(\Rv_j^\prime,\Rv_{j+1}^\prime) \right)
  \right]
 \right\rbrace \,,
\end{equation}
where $\Rv_j^\prime$ collectively denotes the new coordinates at the time slice
$j$. We note that the newly proposed coordinates have the same permutation
vector $P$ of the old ones. The parameter $r_\mathrm{max} \le L/2$ can be tuned
to optimize the sampling efficiency. If the initial bead of one of the polymers
is translated outside of the fundamental cell, we recenter it as discussed
above.

\subsubsection{Redraw}

With this update we redraw the part of a polymer between two fixed beads.
As we have already mentioned, various algorithms with similar efficiencies can
be used for this task, but we find the staging algorithm to be convenient
because it allows us to redraw a segment with an arbitrary number of beads, as
opposed, for example, to the bisection
method~\cite{pollock1984simulation,ceperley1995path} that fixes this number to
powers of $2$. We select a particle index $i_0$, an initial bead $j_0$, and the
number of beads $\Delta j \in [2,j_\mathrm{max}]$ involved in the update,
sampling them from uniform random distributions. The final bead is determined
by the bead index $j_1 =  j_0 + \Delta j$. To avoid clutter in the presentation
we restrict ourselves to present the details for the case $j_1 \le M$, noting
that the generalization to $j_1 > M$ is quite straightforward: one needs to
follow the path to the next particle index $p(i_0)$ and use $(j_1\mod M)$ as
the final bead. Moreover in this case one should follow the path to the next
polymer preserving the length of the links, i.e~momentarily translating the
next polymer to the same cell of the bead
$\rv_{i_0,M}$, and translating it back at the end of the staging procedure.
In what follows we also omit the particle index in the position vectors.
The next step is to propose new coordinates for the $\Delta j - 1$ beads
between $j_0$ and $j_1$ using the so-called L\'evy
construction~\cite{levy1940certains}. This allows us to directly sample the
product of the free particle propagators by rewriting them as
\begin{equation}
 \prod_{j=j_0}^{j_1 -1}
 \rho_\mathrm{free}^\mathrm{sp}(\rv_{j}, \rv_{j+1}, \dtau)
 = \frac{1}{(4\pi\lambda \Delta j \, \dtau)^{D/2}}
 \exp \left[
  -\frac{(\rv_{j_0} - \rv_{j_1})^2}{4\lambda \Delta j \, \dtau}
  \right]
 \prod_{j=j_0+1}^{j_1 -1} \frac{1}{(4\pi\lambda a_j \dtau )^{D/2}} \exp \left[
  - \frac{(\rv_j - \rv_j^*)^2}{4\pi\lambda a_j \dtau } \right]\,,
 \label{eq:Levy}
\end{equation}
where we have defined
\begin{equation}
 \rv_j^* = \frac{\rv_{j_1} + (j_1 - j) \rv_{j-1}}{j_1-j+1}
 \,, \quad
 a_j = \frac{j_1-j}{j_1-j+1} \,.
\end{equation}
The above equation shows that we can use the product of gaussians in
eq.~\eqref{eq:Levy} as the a priori sampling distribution for the redraw move
and sequentially sample the beads from $j_0 + 1$ to $j_1-1$ according to the
conditional (free) probability for picking the point $j$ based on the previous
bead $j-1$ and the final bead $j_1$.
The acceptance probability is computed as
\begin{equation}
 A_R = \min \left\lbrace 1,
 \exp \left[
  \sum_{j=j_0}^{j_1-1} \left(  U(\Rv_j,\Rv_{j+1}) -
  U(\Rv_j^\prime,\Rv_{j+1}^\prime) \right)
  \right]
 \right\rbrace \,,
\end{equation}
The permutation vector $P$ is left unchanged by this update and the simulation
parameter $j_\mathrm{max}$ can be tuned to maximize the sampling efficiency.
It is worth pointing out that, if $j_1>M$, the first bead is involved in the
displacement and, if moved outside of the fundamental cell, we shall recenter
it, as discussed above.

\subsubsection{Open/Close}

The open and close moves are sector-changing updates and are particularly
delicate. With the open move we cut a link, creating two loose extremities and
taking the system from the $Z$-sector of closed paths to the $G$-sector with
one worm. With the close move we bind together the two extremities of the worm,
returning to the $Z$-sector.
Since one move is the opposite of the other, we need to carefully weight the
transition probability in order to achieve the detailed balance. In our
implementation we propose to open or close the worm at the same rate,
independently of the sector the simulation is in, and, only afterwards, we
abort the move if either the open move is called within the $G$-sector or the
close move is called within the $Z$-sector.
The updates that we present below differ from the ones originally introduced in
Refs.~\cite{prokof1998exact,prokof1998worm,prokof2001worm,boninsegni2006worma,boninsegni2006wormb}
and are constructed to be compatible with the periodic boundary conditions,
yielding exact results even for systems where the thermal wavelength is of the
order of the system size, regardless of how many beads are involved in the
updates.
In the rest of this subsection we consistently use the notation $X =
 (Z,P,i_H,\rvc_H,\lbrace \Rv_j \rbrace)$ to refer to the configuration in the
$Z$-sector and the notation $\Xh= (G,P,\ih_H,\rvch_H,\lbrace \Rvh_j \rbrace)$
to refer to the configuration in the $G$-sector. Notice that $X$ and $\Xh$
share the same permutation vector $P$ and that, while the probability
distribution in the $Z$-sector doesn't depend on the head index $i_H$ and the
extra bead $\rvc_H$, they differ from the ones of the state $\Xh$. In the
updates presented below, we embed the coordinate of the extra bead $\rvch_H$
into its representation on $\mathbb{R}^D$ as $\rvh_{\ih_H,M} = \rvch_H +
 \widehat{\wv}_{\ih_H} L$ and sample the new configuration based on
$\rvh_{\ih_H,M}$. \\

The open move consists of the following steps:
\begin{itemize}
 \item Select the particle index $\ih_H$ from a uniform random
       distribution.
 \item Select a time slice $j_0 \le
        j_\mathrm{max}^\mathrm{open}$ from a uniform random distribution, with
       the positive integer
       $j_\mathrm{max}^\mathrm{open} < M$ a tunable parameter.
 \item Propose a new value for the position of the new head $\rvh_{i_H,M}$ by
       displacing the point $\rv_{\ih_H,M} \equiv \rvc_{p(\ih_H),0} +
        \wv_{\ih_H} L$
       by a quantity $\Delta \rv$ uniformly sampled  in the space
       $[-\Delta,\Delta]^D$, with $\Delta < L/2$ an adjustable parameter.
 \item Redraw the portion of the polymer $\ih_H$ going from the bead $j_0+1$ to
       the bead $M-1$ by constructing a free particle path starting at
       $\rv_{\ih_H,j_0}$ and ending after $\Delta j = M - j_0$ steps at
       $\rvch_H$ with
       the staging algorithm described above.
 \item Accept the update with probability
       \begin{equation}
        A_O = \min \left\lbrace 1,
        \frac{C N (2 \Delta)^D}{V}
        \exp \left[
         \sum_{j=j_0}^{M-1} \left(  U(\Rv_j,\Rv_{j+1}) -  U(\Rvh_j,\Rvh_{j+1})
         \right)
         \right]
        \frac{\rho_\mathrm{free}^\mathrm{sp}(\rvh_{\ih_H,j_0}, \rvh_{\ih_H,M},
         \Delta j \, \dtau)}
        {\rho_\mathrm{free}^\mathrm{sp}(\rv_{\ih_H,j_0}, \rv_{\ih_H,M},
         \Delta j \, \dtau)}
        \right\rbrace \,.
        \label{eq:A_open}
       \end{equation}
\end{itemize}

The close move provides the detailed balanced complement to the open move. In
explaining the steps for the close move we remind the reader that as per the
open move detailed above, the position vectors $\lbrace \Rv_j \rbrace$ and
$\lbrace \Rvh_j \rbrace$ only differ at particle $\ih_H$ for the bead indices
going from $j_0+1$ to $M$.
The close move consists of the following steps:
\begin{itemize}
 \item Identify the particle indices $\ih_H$ and $\ih_T=p(\ih_H)$ corresponding
       to the head and the tail of the worm, respectively.
 \item Select a time slice $j_0 \le j_\mathrm{max}^\mathrm{open}$ from a
       uniform random distribution.
 \item Find the periodic image $\rvh_{T} = \rvch_{\ih_T,0} + \wv_{\ih_H} L $ of
       the first bead of the tail $\rvch_{\ih_T,0}$ that is the nearest to the
       head bead $\rvh_{\ih_H,M}$ and check whether their difference
       $\rvh_{\ih_H,M} -
        \rvh_{T}$ is
       within $[-\Delta,\Delta]$ in every direction. If that is the case set
       $\rv_{\ih_H,M} = \rvh_T$, otherwise abort the update.
 \item Redraw the portion of the polymer $\ih_H$ going from the bead $j_0+1$ to
       the bead $M-1$ by constructing a free particle path starting at
       $\rv_{\ih_H,j_0}$ and ending after $\Delta j = M - j_0$ steps at
       $\rv_{\ih_H,M}$ with the staging algorithm.

 \item Accept the update with probability
       \begin{equation}
        A_C = \min \left\lbrace 1,
        \frac{V}{C N (2\Delta)^D}
        \exp \left[
         \sum_{j=j_0}^{M-1} \left(  U(\Rvh_j,\Rvh_{j+1}) -  U(\Rv_j,\Rv_{j+1})
         \right)
         \right]
        \frac{
         \rho_\mathrm{free}^\mathrm{sp}(\rv_{\ih_H,j_0}, \rv_{\ih_H,M}, \Delta
         j \, \dtau)} {\rho_\mathrm{free}^\mathrm{sp}(\rvh_{\ih_H,j_0},
         \rvh_{\ih_H,M},
         \Delta j \, \dtau)}
        \right\rbrace \,.
        \label{eq:A_close}
       \end{equation}
\end{itemize}
As a last step, one should randomly select the particle index $i_H$ and the
value of $\rvc_H$ inside the fundamental cell, but, since the PDF in the
$Z$-sector does not depend on them, one can actually forget to update their
value in the simulation.\\

Given the delicate nature of the present moves it is instructive to explicitly
derive the acceptance probability in eqs.~\eqref{eq:A_open}
and~\eqref{eq:A_close} to highlight the different elements that make the
sector-changing moves possible. We first write the ratio of the PDFs,
\begin{equation}
 \frac{\pi_W(\Xh)}{\pi_W(X)} = C N
 \frac{\prod_{j=j_0}^{M-1}
  \rho_\mathrm{free}^\mathrm{sp}(\rvh_{\ih_H,j},\rvh_{\ih_H,j+1},\dtau)}
 {\prod_{j=j_0}^{M-1}\rho_\mathrm{free}^\mathrm{sp}(\rv_{\ih_H,j},\rv_{\ih_H,j+1},\dtau)}
 \exp\left[
  \sum_{j=j_0}^{M-1} \left(  U(\Rv_j,\Rv_{j+1}) -  U(\Rvh_j,\Rvh_{j+1}) \right)
  \right]\,,
\end{equation}
and then we evaluate the transition probabilities. Since the choice of the
particle indices $i_H$ and $\ih_H$ and the choice of the time slice $j_0$ are
based on uniform distributions and proceed identically for
the open and
close moves, we don't report the corresponding factors below since they
simplify in the ratio and are not relevant for the discussion. For the open
move the procedure explained above gives
\begin{equation}
 T(X,\Xh) =
 \frac{1}{(2 \Delta)^D}
 \prod_{j=j_0+1}^{M -1} \frac{1}{(4\pi\lambda a_j \dtau )^{D/2}} \exp \left[ -
  \frac{(\rvh_{\ih_H,j} - \rvh_{\ih_H,j}^*)^2}{4\pi\lambda a_j \dtau }
  \right]\,,
 \label{eq:T_open}
\end{equation}
with $\rvh_{\ih_H,j}^*$ and $a_j$ obtained with the Levy construction as in
eq.~\eqref{eq:Levy}. For the close move we have instead
\begin{equation}
 T(\Xh,X) = \frac{1}{V}
 \prod_{j=j_0+1}^{M -1} \frac{1}{(4\pi\lambda a_j \dtau )^{D/2}} \exp \left[ -
  \frac{(\rv_{\ih_H,j} - \rv_{\ih_H,j}^*)^2}{4\pi\lambda a_j \dtau } \right]\,,
 \label{eq:T_close}
\end{equation}
with $\rv_{\ih_H,j}^*$ obtained with the Levy construction and with the factor
$1/V$ coming from uniformly sampling the vector $\rvc_H$. Putting these
equations together and using the identity~\eqref{eq:Levy} one gets the
acceptance probabilities in eqs.~\eqref{eq:A_open} and~\eqref{eq:A_close}.
As stated at the beginning of the subsection, the open/close updates introduced
here are compatible with the periodic boundary conditions and the parameter
$j_\mathrm{max}^\mathrm{open}$ can be freely chosen to maximize the sampling
efficiency.
In contrast to the original algorithm, where one faces the ambiguity in
the choice of the image of the tail $\rv_T$ when closing the
polymer---resulting in a violation of the detailed balance condition when the
thermal wavelength is of the order of the size of the system---the updates
presented here are always perfectly balanced. This is achieved by a different
sampling choice for $\rvh_{\ih_H,M}$, with the cutoff $\Delta < L/2$ removing
said ambiguity. A
convenient choice for the parameter $\Delta$ is to make it dependent on the
choice of $j_0$ as
\begin{equation}
 \Delta = \min ( \sqrt{2\lambda (M-j_0) \, \dtau}, L/2)\,.
\end{equation}

\subsubsection{Swap}

The swap update allows one to sample the permutations in a very efficient way
by connecting the worm head to a near  polymer. It requires the worm to be
present, hence it is performed only in the $G$-sector, meaning that it must be
aborted if proposed in the $Z$-sector. We denote with $i_H$ the particle index
associated with the worm's head. We first select from a uniform random
distribution a time slice $j_P \le  j_\mathrm{max}^\mathrm{swap}$ with
$j_\mathrm{max}^\mathrm{swap}<M$, which
will act as pivot. We first compute the free propagators
\begin{equation}
 \Pi_P(i) = \rho_\mathrm{free}^\mathrm{sp} (\rvc_H,\rv_{i,j_P}, j_P \dtau)\,,
 \label{eq:Pi}
\end{equation}
for every particle $i$, and we then use tower sampling to select a particle
index $i_0$ with probability $\Pi_P(i)/\Sigma_P$, where the normalization
factor is given by
\begin{equation}
 \Sigma_P = \sum_{i=1}^{N} \Pi_P(i)\,.
\end{equation}
Before proceeding further we need to test the particle $i_0$. If it corresponds
to the tail of the worm $i_T$ we abort the move and reject the update. This is
necessary to prevent the worm from closing on itself, and hence disappearing.
If that is not the case we proceed by computing the normalization factor for
the inverse process
\begin{equation}
 \Sigma_0 = \sum_{i=1}^{N} \rho_\mathrm{free}^\mathrm{sp}
 (\rvc_{i_0,0},\rv_{i,j_P}, j_P \dtau)\,,
 \label{eq:S0}
\end{equation}
which is necessary to ensure the detailed balance.
Notice that the free propagators in eqs.~\eqref{eq:Pi} and~\eqref{eq:S0} are
computed using the simulation coordinates, without invoking periodic boundary
conditions.
We then cut the polymer $i_0$ and we bind its first bead to the
head of the
worm by setting $\rvc_{i_0,0}'=\rvc_H$. We finally construct a free particle
path $\rv_{i_0,1}', \dots, \rv_{i_0,j_P-1}'$ with the staging algorithm.
The acceptance probability is computed as
\begin{equation}
 A_{SW} = \min \left\lbrace 1, \frac{\Sigma_P}{\Sigma_0}
 \exp \left[
  \sum_{j=0}^{j_P-1} \left(  U(\Rv_j,\Rv_{j+1}) -
  U(\Rv_j^\prime,\Rv_{j+1}^\prime) \right)
  \right]
 \right\rbrace \,,
\end{equation}
Notice that this update changes the permutation $P$ to $P'$ such that $p'(i_H)
 = i_0$, and the new worm's head $i_H'$ is instead identified as the particle
that was in permutation with $i_0$ before the update, i.e.~we set $i_H'=i^*$
where $i^*$ is such that $P(i^*)=i_0$. The tail index remains unchanged.

\subsubsection{Move head}

This and the next move are performed only in the $G$-sector, when a worm is
present. They must be aborted if proposed in the  $Z$-sector. With \emph{move
 head} we redraw the last beads of the worm. We first select a starting time
slice $j_0$ and then sample the new worm's head bead $\rv_{i_H,M}'$, from the
free particle propagator
\begin{equation}
 \rho_\mathrm{free}^\mathrm{sp}(\rv_{i_H,j_0}, \rv_{i_H,M}', \Delta j \, \dtau)
 =
 (4\pi\lambda \Delta j\, \dtau )^{-D/2} \exp \left[
  -\frac{ (\rv_{i_H,j_0} -\rv_{i_H,M}')^2}{4\lambda\Delta j\,\dtau}\right] \,,
\end{equation}
where $i_H$ is the particle index of the worm's head and $\Delta j = M-j_0$.
We then construct a free path $\rv_{i_H,j_0+1}', \dots, \rv_{i_H,M-1}'$ with
the staging algorithm. We accept the update with probability
\begin{equation}
 A_H = \min \left\lbrace 1,
 \exp \left[
  \sum_{j=j_0}^{M-1} \left(  U(\Rv_j,\Rv_{j+1}) -
  U(\Rv_j^\prime,\Rv_{j+1}^\prime) \right)
  \right]
 \right\rbrace \,.
\end{equation}

\begin{figure}[t]
 %\vspace*{15pt}
 \includegraphics[width=0.47\linewidth]{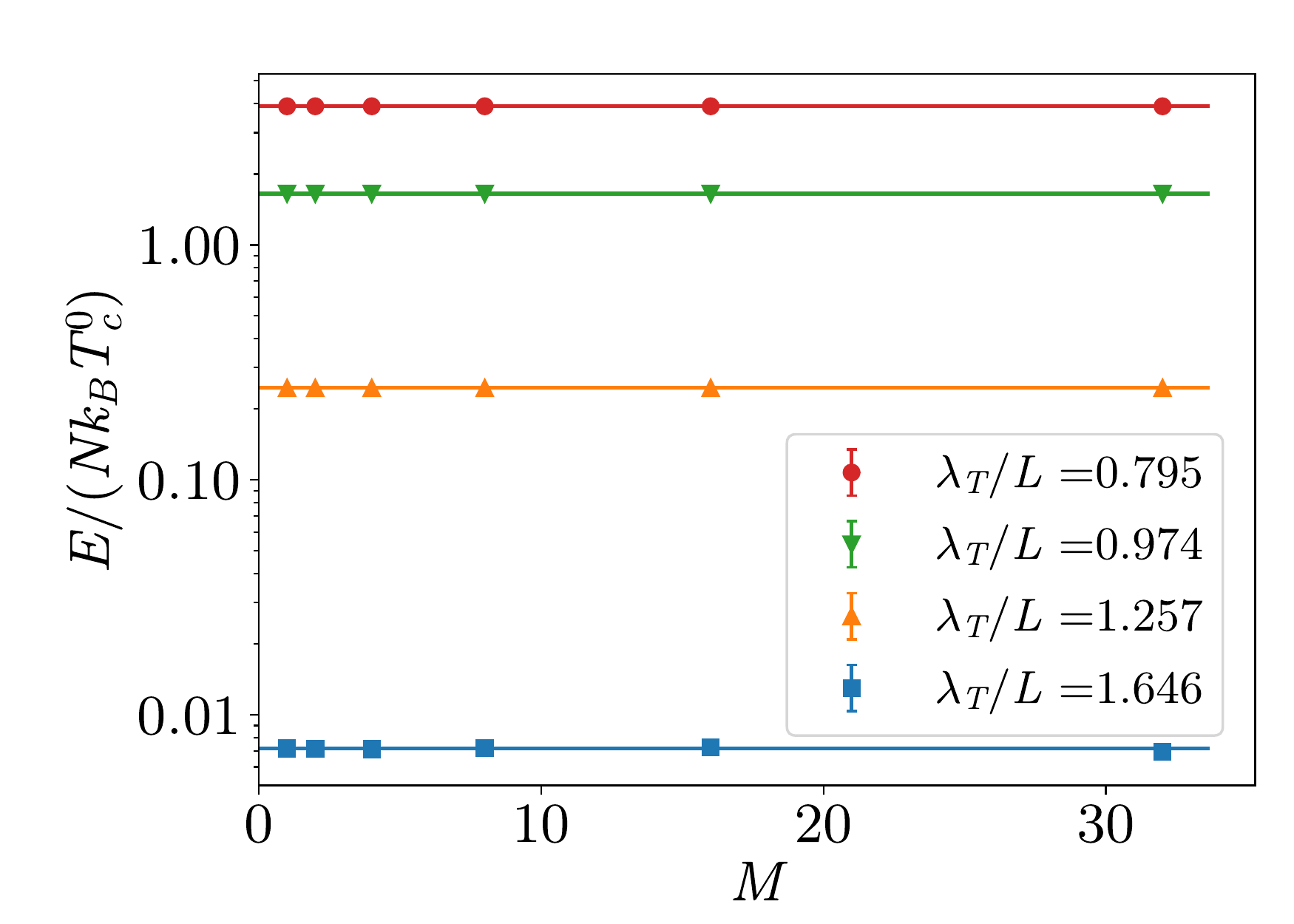}~~
 \includegraphics[width=0.47\linewidth]{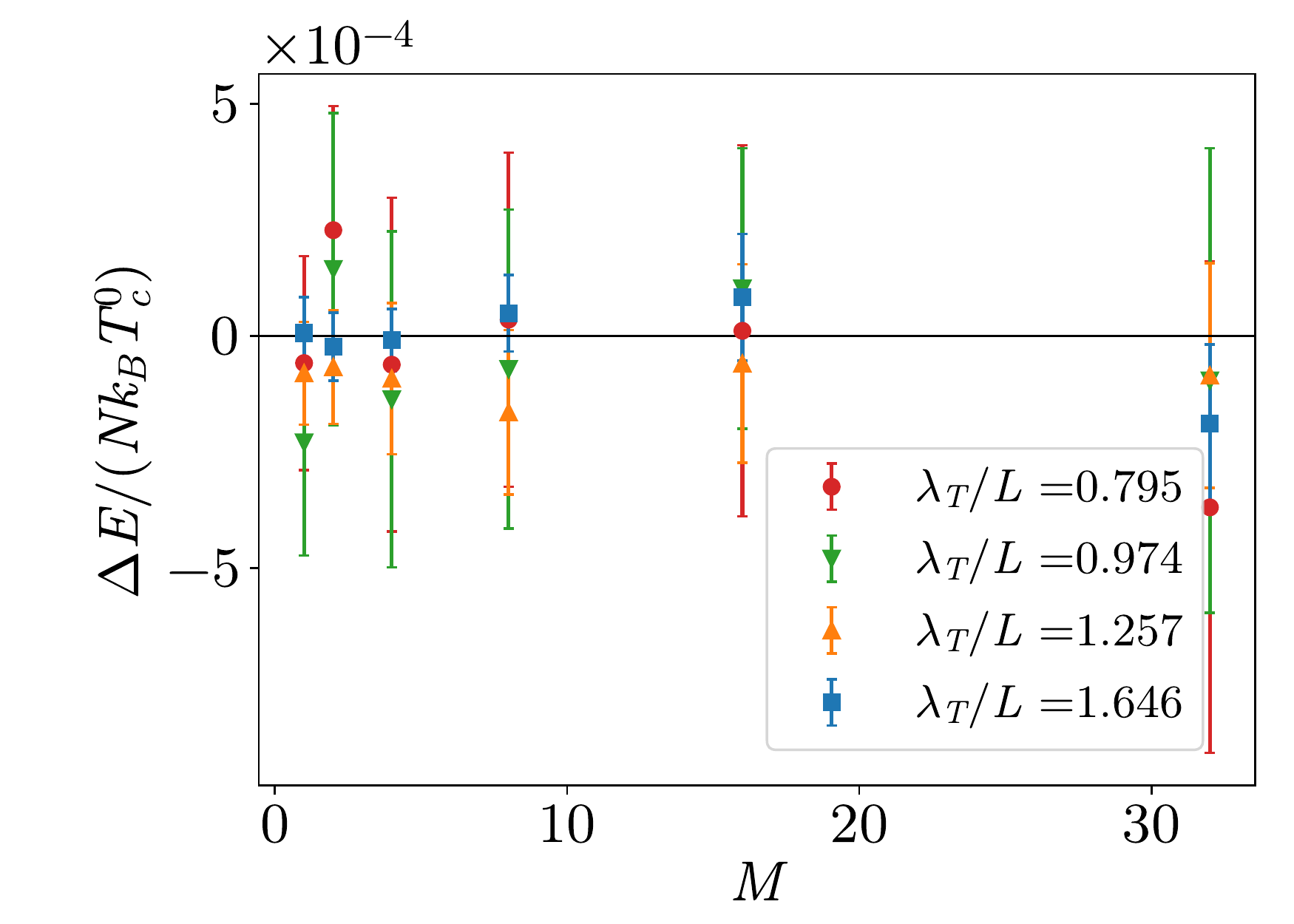}
 \caption{Results for the $N=1$ system at different values of $\lambda_T/L$.
  \emph{Left:} Internal energy $E$, in units of $N k_B T_c^0$, with $T_c^0$
  being
  the critical temperature defined in eq.~\eqref{eq:Tc0}. The values obtained
  at
  different number of beads $M$ are compared with the exact results (horizontal
  lines). \emph{Right:} Differences with the exact values.}
 \label{fig:1}
\end{figure}

\subsubsection{Move tail}

Similarly, with \emph{move tail} we redraw the first beads of the worm by
selecting a time slice $j_0$ and generating a proposal $\rv_T'$ for the new
tail bead of the worm $\rvc_{i_T,0}$ according to the distribution
\begin{equation}
 \rho_\mathrm{free}^\mathrm{sp}( \rv_T', \rv_{i_T,j_0}, \Delta j \, \dtau) =
 (4\pi\lambda \Delta j\, \dtau )^{-D/2} \exp \left[
  -\frac{ (\rv_T' - \rv_{i_T,j_0})^2}{4\lambda\Delta j\,\dtau}\right] \,,
\end{equation}
where $i_T$ is the particle index of the worm's tail and $\Delta j = M-j_0$. We
set $\rvc_{i_T,0}=\rv_T'$ and then construct a free path $\rv_{i_T,1}', \dots,
 \rv_{i_T,j_0-1}'$ with the staging algorithm.
We accept the update with probability
\begin{equation}
 A_T = \min \left\lbrace 1,
 \exp \left[
  \sum_{j=j_0}^{M-1} \left(  U(\Rv_j,\Rv_{j+1}) -
  U(\Rv_j^\prime,\Rv_{j+1}^\prime) \right)
  \right]
 \right\rbrace \,.
\end{equation}
If the coordinate $\rv_T'$ falls outside of the fundamental cell, we recenter
the tail polymer as discussed above.

\section{Non-interacting Bose gas}
\label{sec:III}

\begin{figure}[t]
 \includegraphics[width=0.47\linewidth]{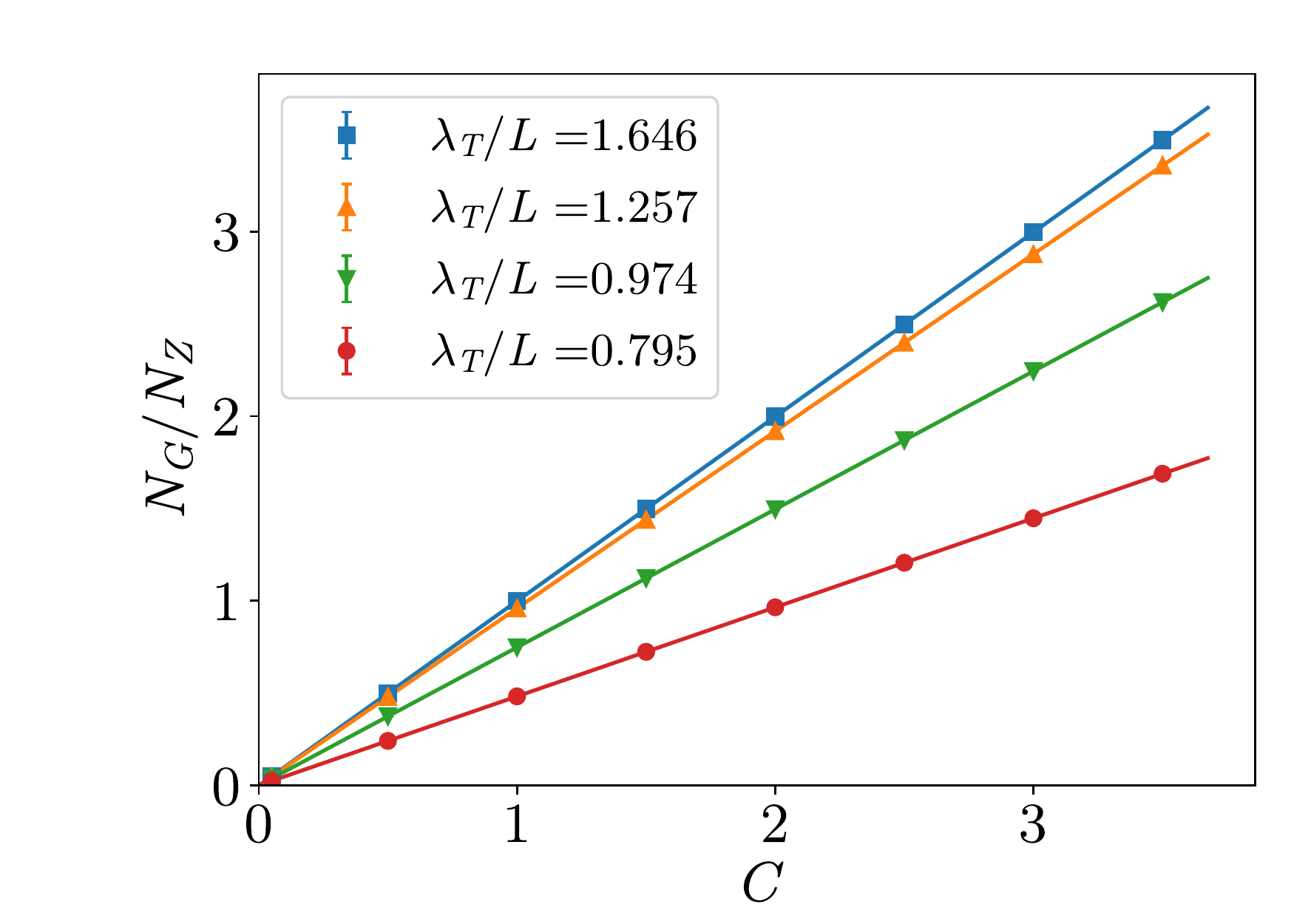}~~
 \includegraphics[width=0.47\linewidth]{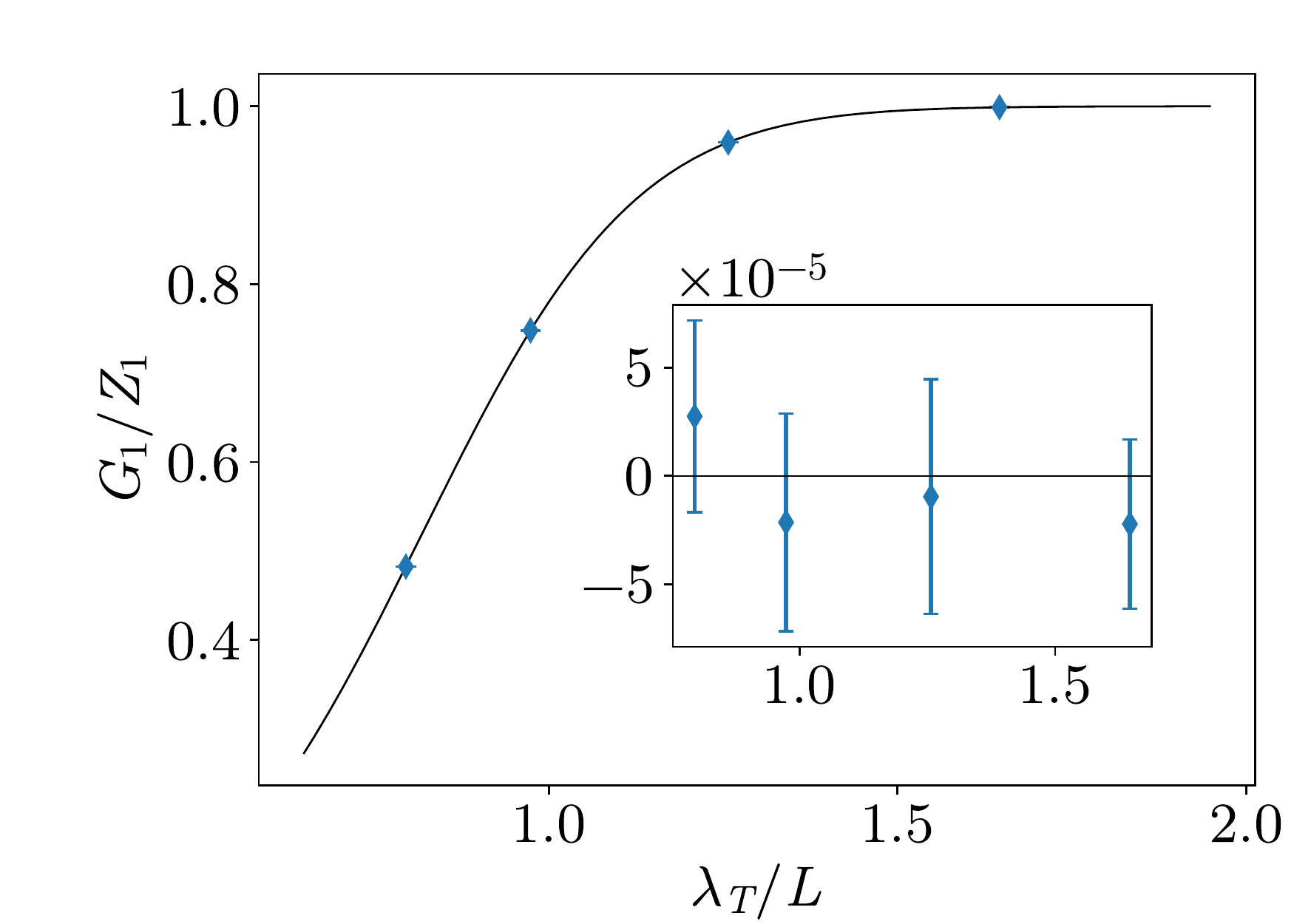}
 \caption{Results for the $N=1$ system at different values of $\lambda_T/L$.
  \emph{Left:} check of the proportionality in eq.~\eqref{eq:NG_NZ}, the lines
  are one-parameter fits. \emph{Right:} the proportionality coefficients
  extracted from the fits (diamonds) are compared with the exact value (solid
  line). Inset: difference between the coefficients and the exact value.}
 \label{fig:2}
\end{figure}

When developing a code, one should have precise benchmarks aimed at
validating the various aspects of the algorithm. In this section we start with
tests for non-interacting systems, where exact results are known, and we will
consider interacting systems in the next section. From now on we specialize to
the $D=3$ case.
Textbook treatments of the non-interacting Bose gas model usually consider the
system in the grand-canonical ensemble and in the thermodynamic limit. However,
exact solutions are available also for a fixed number of particles in a box
with periodic boundary conditions. In particular, we are interested in the
calculation of the internal energy
\begin{equation}
 E=-\frac{1}{Z_N}\frac{\partial Z_N}{\partial\beta} \;.
 \label{eq:ideal1}
\end{equation}
The partition function $Z_N(\beta)$ for $N$ particles at inverse temperature
$\beta$ can be calculated using the recursion
formula~\cite{doi:10.1063/1.464180,krauth06}
\begin{equation}
 Z_N(\beta)=\frac{1}{N} \sum_{k=1}^N z(k\beta)Z_{N-k}(\beta) \;,
 \label{eq:ideal2}
\end{equation}
involving the partition functions of $N-k$ particles at the same inverse
temperature $\beta$ with the starting value $Z_0(\beta)=1$ and the partition
function of a single particle $z(k\beta)$ at the multiple inverse
temperature $k\beta$. The
latter partition function for our system with periodic boundary conditions is
defined as
\begin{equation}
 z(\beta)=\sum_{n_x,n_y,n_z}e^{-\beta \varepsilon(n_x,n_y,n_z)} \;,
\end{equation}
where $\varepsilon(n_x,n_y,n_z)$ are the single-particle energies labeled by
the integers $n_{x,y,z}=0,\pm1,\pm2,\dots$
\begin{equation}
 \varepsilon(n_x,n_y,n_z) = \lambda \left(\frac{2\pi}{L}\right)^2
 (n_x^2+n_y^2+n_z^2) \;.
\end{equation}
The calculation of the internal energy $E$ in Eq.~\eqref{eq:ideal1} can be
carried
out recursively from the derivatives with respect to $\beta$ using result
\eqref{eq:ideal2}.\\

In the following we perform PIMC calculations of the internal energy of $N$
non-interacting bosons at different temperatures and we directly compare the
results with the exact value obtained from Eq.~\eqref{eq:ideal1}. First we
compare
for the case $N=1$, where no swap moves are involved in the PIMC simulation but
it provides a stringent benchmark for all the other moves. Then we move to
$N=2$ in order to test the correct implementation also of the swap moves.
Finally we consider the case $N\gg1$ and the approach to the thermodynamic
limit.

\subsection{Benchmarks}

We carry out a first non-trivial check on the worm algorithm by
considering a single particle in a regime where the thermal wavelength
$\lambda_T$ is of the order of the side of the box $L$, thus testing the
compatibility of the updates (except swap) with the periodic boundary
conditions. In fig.~\ref{fig:1} we present the results for the internal energy
$E$ for different values of the ratio $\lambda_T/L$ and compare them with the
exact results.
We performed the test varying the total number of beads $M$ while keeping
unconstrained the simulation parameters controlling the portion of the polymers
involved in the moves, namely setting $j_\mathrm{max}$ and
$j_\mathrm{max}^\mathrm{open}$ to the maximum value $M$.
Since, in the absence of the interaction, the algorithm is exact for any number
of beads, we recover the exact result even for $M=1$.\\

We then turn a closer look to the open/close update, verifying the
proportionality between the ratio $N_G/N_Z$ and the parameter $C$, as expressed
in eq.~\eqref{eq:NG_NZ}. In the left panel of fig.~\ref{fig:2} we report the
ratios obtained at different values of $C$ together with one-parameter fits. In
the right panel, the fitted coefficient is in turn compared with the exact
value of $G_1/Z_1$ given by
\begin{equation}
 \frac{G_1}{Z_1}
 = \left[
  \vartheta_3 \left(0, e^{-\pi \frac{\lambda_T^2}{L^2}}\right)
  \right]^{-3}\,,
\end{equation}
where $\vartheta_3(z,q)$ is the theta function with rational
characteristic $3$ (see e.g.~chapter 16 of ref.~\cite{Abramowitz1965}). As one
can see from the figure, there is perfect agreement between the
PIMC points and the expected results, with the inset showing the difference
between them.\\

\begin{figure}[t]
 \includegraphics[width=0.47\linewidth]{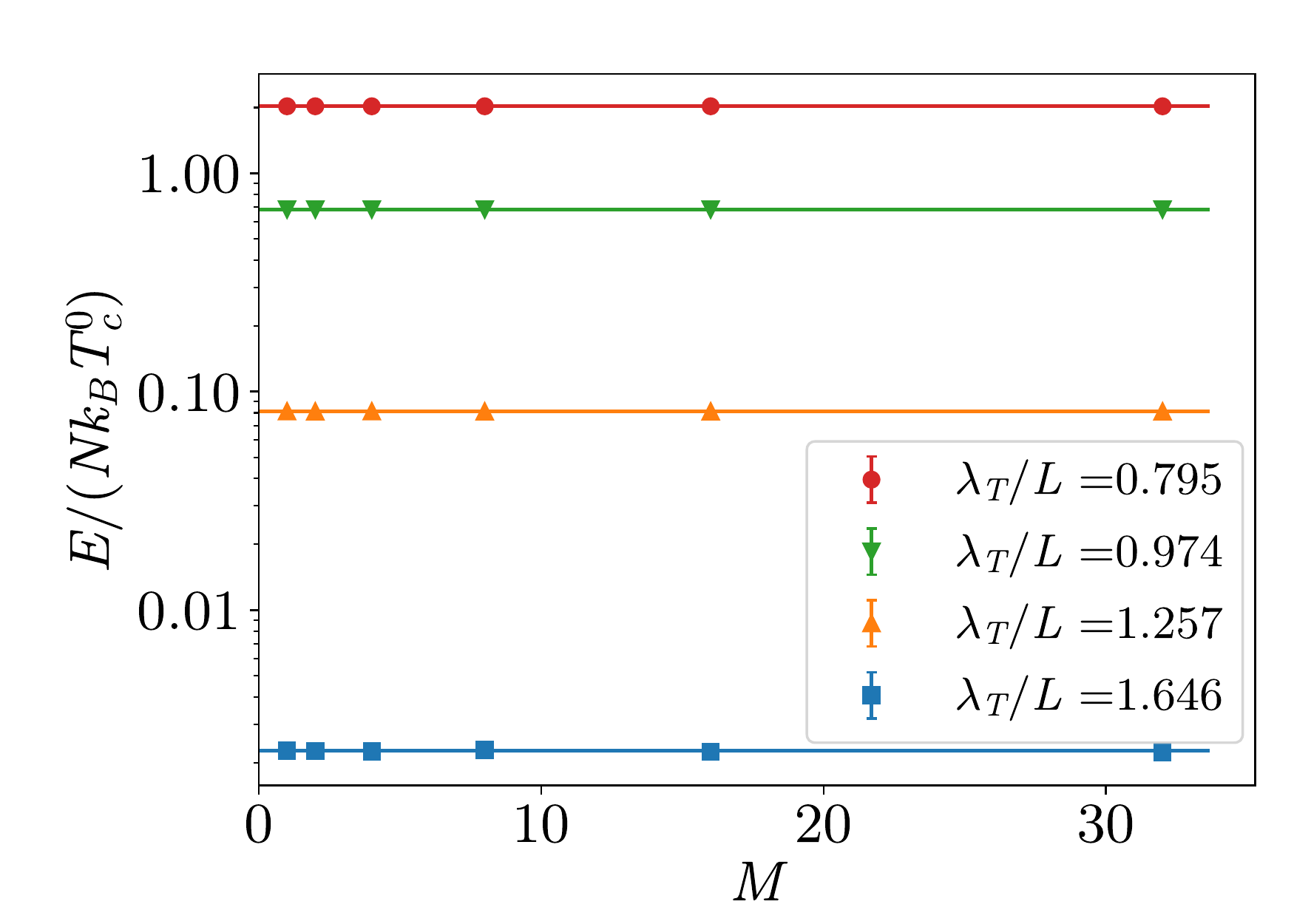}~~
 \includegraphics[width=0.47\linewidth]{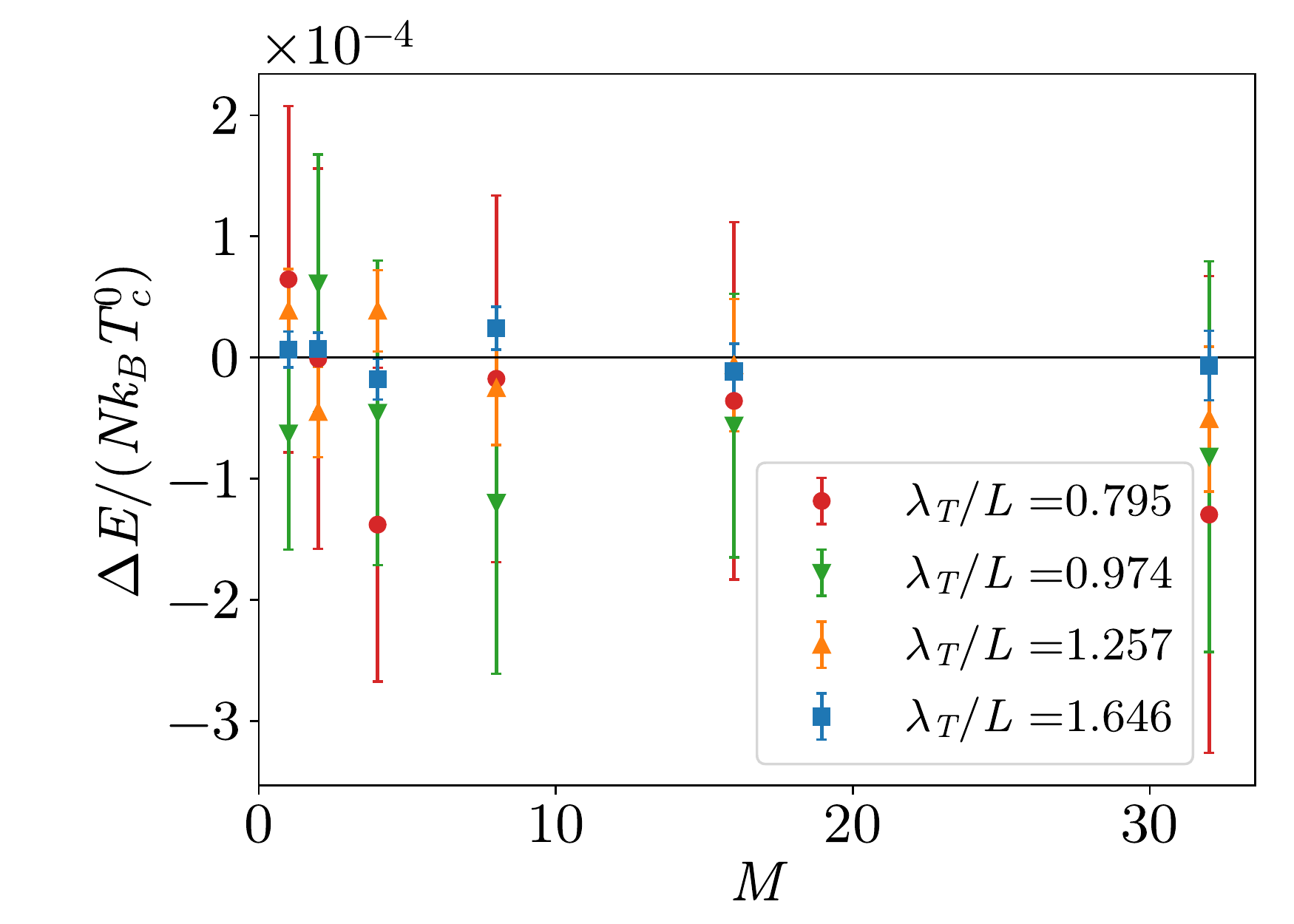}
 \caption{Results for the $N=2$ system at different values of $\lambda_T/L$.
  \emph{Left:} Internal energy $E$ computed with different number of beads $M$
  and compared with the exact results (horizontal lines). As in
  fig.~\ref{fig:1} the energy scale is defined in eq.~\eqref{eq:Tc0}.
  \emph{Right:} Differences with the exact values.}
 \label{fig:3}
\end{figure}

We now move to the $N=2$ system where the swap update makes its first
appearance and we check the results for different values of $\lambda_T/L$
and various values of $M$. As before, we used unconstrained parameters
$j_\mathrm{max} = j_\mathrm{max}^\mathrm{open} = j_\mathrm{max}^\mathrm{swap} =
 M$, allowing updates involving a whole polymer. We report the results in
fig.~\ref{fig:3}, where we show that we recover the exact result for any number
of beads used, meaning that the implementation of the swap is fully
compatible with periodic boundary conditions.\\

\begin{figure}[t]
 \includegraphics[width=0.47\linewidth]{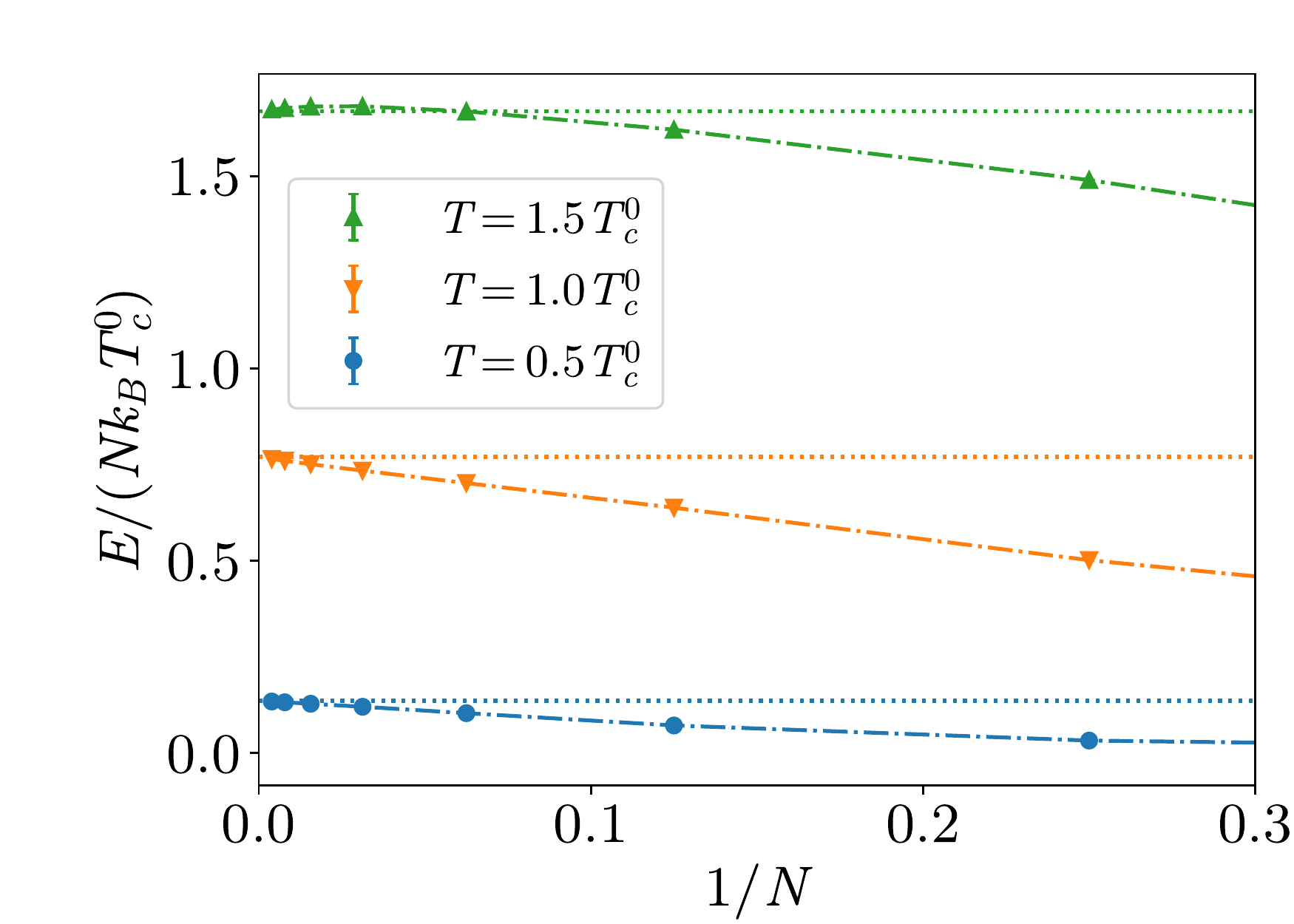}~~
 \includegraphics[width=0.47\linewidth]{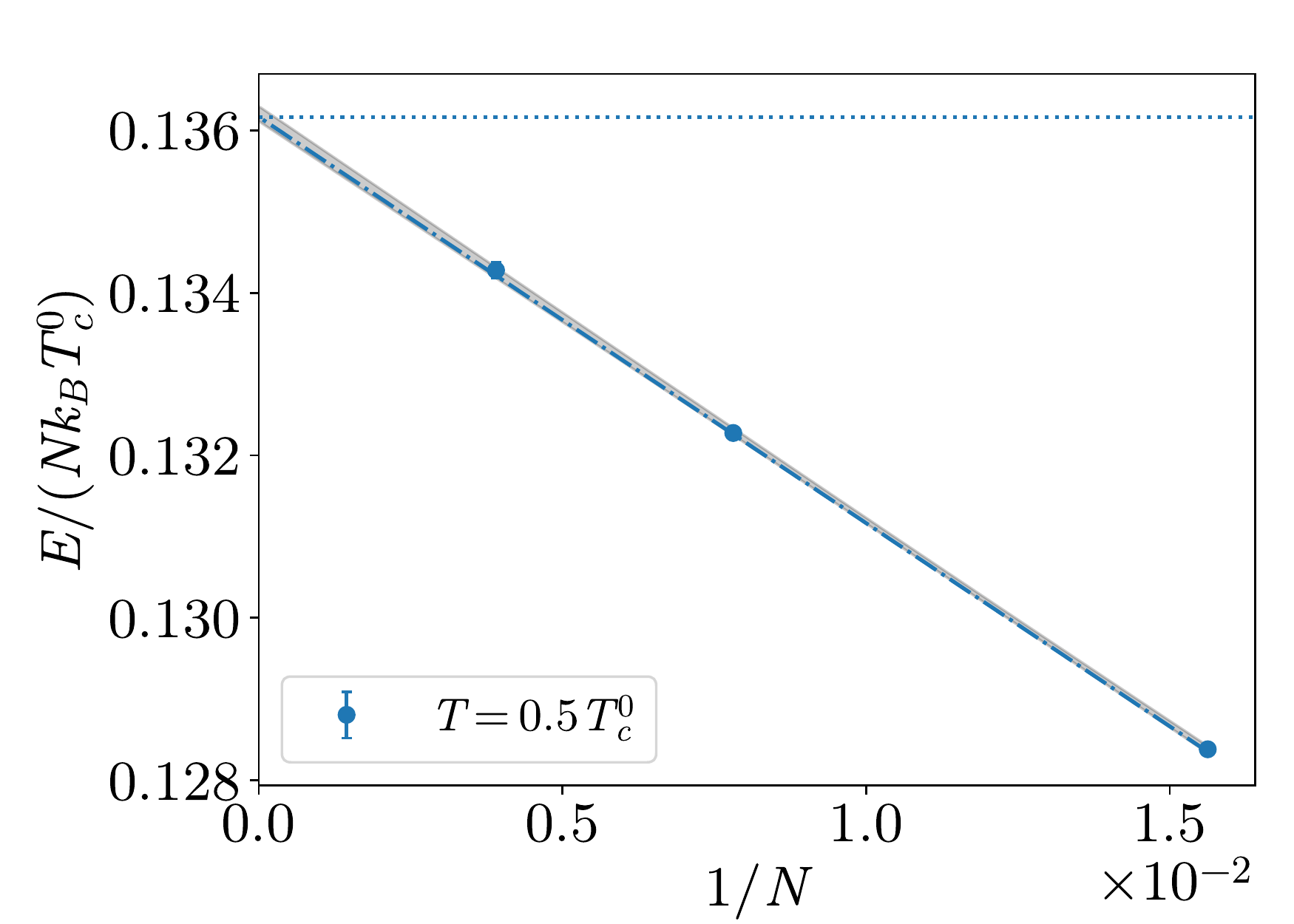}
 \caption{Internal energy $E$ for $N$-particle systems at fixed temperature.
  \emph{Left:} Worm algorithm results compared with the exact values at fixed
  number of particles (dot-dashed lines) and with the exact thermodynamic limit
  (horizontal dotted lines). Error bars, being of the order of $10^{-4}$, are
  not visible on this scale and are perfectly compatible with the exact
  results.
  \emph{Right:} Zoom on the $T=0.5T_c^0$ data, with a linear fit extrapolation
  to the thermodynamic limit. The gray band covers the one-sigma region of the
  fit.}\label{fig:4}
\end{figure}

The one and two-particle systems provide clean---and to some extent
independent---tests for the whole set of updates in the regime where the effect
of periodic boundary conditions is the largest. We now show that the exactness
of the algorithm is preserved when taking larger and larger numbers of
particles. For this purpose we vary $N$ while keeping the temperature $T$
fixed. In fig.~\ref{fig:4} we show the results for the internal energy for
three values of the temperature, expressed in units of the critical temperature
$T_c^0$ defined as
\begin{equation}
 k_B T_c^0  = 4\pi\lambda \left(\frac{n}{\zeta(3/2)}\right)^{2/3} \;,
 \label{eq:Tc0}
\end{equation}
where $n = N/V$ is the number density. For every value of $N$ the worm
algorithm gives results which are in agreement with the exact values computed
from eq.~\eqref{eq:ideal2} and shown in the left panel of fig.~\ref{fig:4} with
the dot-dashed lines. Notice that the error bars are of the order of $10^{-4}$
and are completely hidden by the symbols. The horizontal dotted lines represent
the value of the internal energy in the thermodynamic limit. In the right panel
of fig.~\ref{fig:4} we show a linear fit in $1/N$ that extrapolates to the
thermodynamic limit the three largest system sizes at $T=0.5 T_c^0$. The gray
band covers the $1$-sigma region of the linear fit and perfectly recovers the
$N \to\infty$ result shown by the horizontal dotted line.

\section{Hard-spheres Bose gas}
\label{sec:IV}

In this section we consider interacting systems described by the following
microscopic Hamiltonian with two-body interactions
\begin{equation}
 H = - \lambda \sum_{i=1}^N\nabla_i^2+\sum_{i<k}v(|\rv_i-\rv_k|)\,,
\end{equation}
where $m$ is the mass of the $N$ identical bosons and  $\rv_i$ indicates the
particle position vector. For pedagogical reasons we use a simple interatomic
potential corresponding to a purely repulsive hard core, without any attractive
tail to avoid the occurrence of possible cluster bound states. More precisely,
we use the hard-sphere (HS) model defined as
\begin{equation}
 v(r) =
 \begin{cases}
  +\infty & (r<a) \,, \\
  0       & (r>a) \,,
 \end{cases}
\end{equation}
in terms of the HS diameter $a$. Besides the degeneracy parameter
$n\lambda_T^3$, only one extra parameter, the gas parameter $na^3$, is needed
to
fully characterize the many-body physics of the model. Equilibrium states
of the system correspond either to a gas or to a solid, the latter requiring
that the gas parameter be large enough and the temperature small
enough~\cite{PhysRevA.3.776,PhysRevA.9.2178}. Furthermore, in the limit of a
small gas parameter, the HS model fully captures the universal behavior of
dilute gases in terms of the $s$-wave scattering length, which coincides with
the HS diameter $a$. A careful study of the thermodynamics of the HS model in
the dilute regime has been carried out in Ref.~\cite{PhysRevA.105.013325}.
Here, for illustrative reasons, we consider the HS gas at a much higher density
$na^3=0.1$, which is not so far from the corresponding density of liquid $^4$He
and where the issues of convergence with the number of beads are more relevant.\\

A convenient approximation scheme for the high temperature density matrix
entering the PIMC algorithm is the pair-product ansatz~\cite{ceperley1995path}
\begin{equation}
 \rho(\Rv,\Rv^\prime,\dtau) =
 \prod_{i=1}^N
 \rho_\mathrm{free}^\mathrm{sp} (\rv_i, \rv_i^\prime, \dtau)
 \prod_{i<k}
 \frac{\rho_{rel}(\rv_{ik},\rv_{ik}^\prime,\dtau)}
 {\rho_{rel}^0(\rv_{ik},\rv_{ik}^\prime,\dtau)} \;.
 \label{eq:pair_prod}
\end{equation}
In the above equation $\rho_\mathrm{free}^\mathrm{sp}$ is the
single-particle ideal-gas density matrix defined in eq.~\eqref{eq:rho_free}
and $\rho_{rel}$ is the two-body density matrix of the
interacting system, which depends on
the relative coordinates  $\rv_{ik}=\rv_i-\rv_k$ and
$\rv_{ik}^\prime=\rv_i^\prime-\rv_k^\prime$,
divided by the corresponding ideal-gas term
\begin{equation}
 \rho_{rel}^0(\rv_{ik},\rv_{ik}^\prime,\dtau)=
 \left(8\pi \lambda \dtau \right)^{-3/2}
 \exp\left[-\frac{(\rv_{ik}-\rv_{ik}^\prime)^2}{8\lambda \dtau}\right] \;.
\end{equation}
The advantage of the decomposition in eq.~\eqref{eq:pair_prod} is that
the two-body density matrix at the inverse temperature $\dtau$,
$\rho_{rel}(\rv,\rv^\prime,\dtau)$, can be calculated exactly for a
given potential $V(r)$, thereby solving by construction the two-body problem.
This is the most effective strategy when the system is dilute, but can also be
pursued at higher density. For the HS potential a simple and remarkably
accurate analytical approximation of the high-energy two-body density matrix is
due to Cao and	Berne~\cite{doi:10.1063/1.463076}. For $r>a$ and $r^\prime>a$,
the result is given by
\begin{eqnarray}
 \frac{\rho_{rel}(\rv,\rv^\prime,\dtau)}
 {\rho_{rel}^0(\rv,\rv^\prime,\dtau)}=
 1 -\frac{a(r+r^\prime)-a^2}{rr^\prime}
 e^{-[rr^\prime +a^2-a(r+r^\prime)](1+\cos\theta)/(4\lambda \dtau)} \;,
\end{eqnarray}
where $\theta$ is the angle between the directions of $\bf{r}$ and
$\bf{r}^\prime$, while it vanishes when either $r$ or $r^\prime$ are smaller
than $a$. An important remark concerning the Cao--Berne approximation is that it
correctly describes the scattering of hard spheres at high energy (small
$\delta_\tau$), and it exactly accounts for the $s$-wave term in the partial wave
expansion, which is the dominant contribution at low energy.
Within the Cao--Berne approximation, the interaction energy of hard spheres
evaluated at two subsequent beads $j$ and $j+1$ can finally be written as
(setting $\Rv=\Rv_{j}$ and $\Rv^\prime=\Rv_{j+1}$)
\begin{equation}
 U(\Rv, \Rv^\prime) = - \sum_{i<k} \log \bigg(
 \dfrac{\rho_{rel}(\rv_{ik},\rv_{ik}^\prime,\dtau)}
  {\rho_{rel}^0(\rv_{ik},\rv_{ik}^\prime,\dtau)} \bigg) \,.
 \label{eq:U_CaoBerne}
\end{equation}
In the following subsections we investigate the convergence with the number of
beads for different number of particles and different temperatures at the
density $na^3=0.1$. Finally we investigate the approach to the thermodynamic
limit for increasing values of $N$ at a specific temperature.

\begin{figure}[t]
	\includegraphics[width=0.47\linewidth]{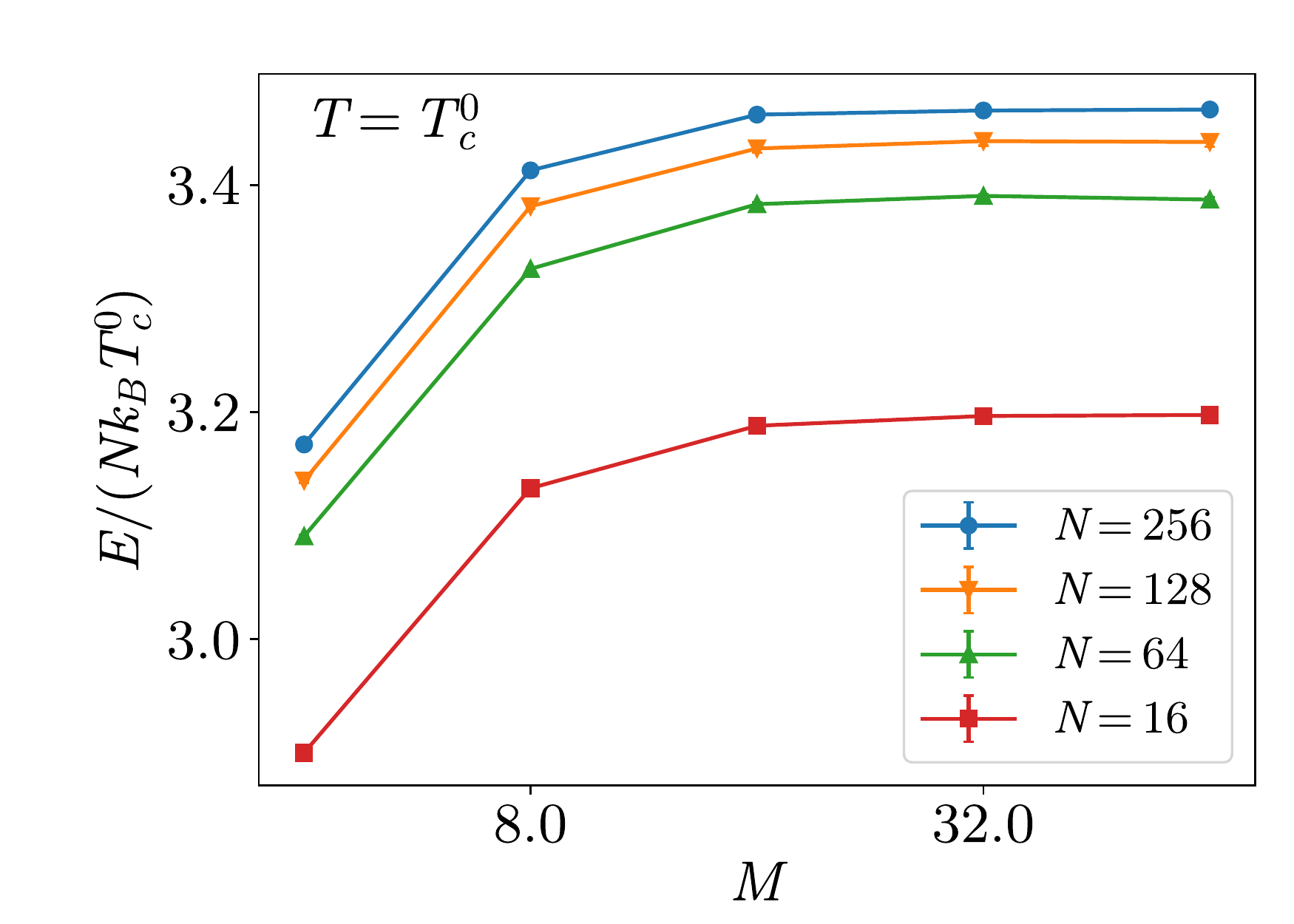}~~
	\includegraphics[width=0.47\linewidth]{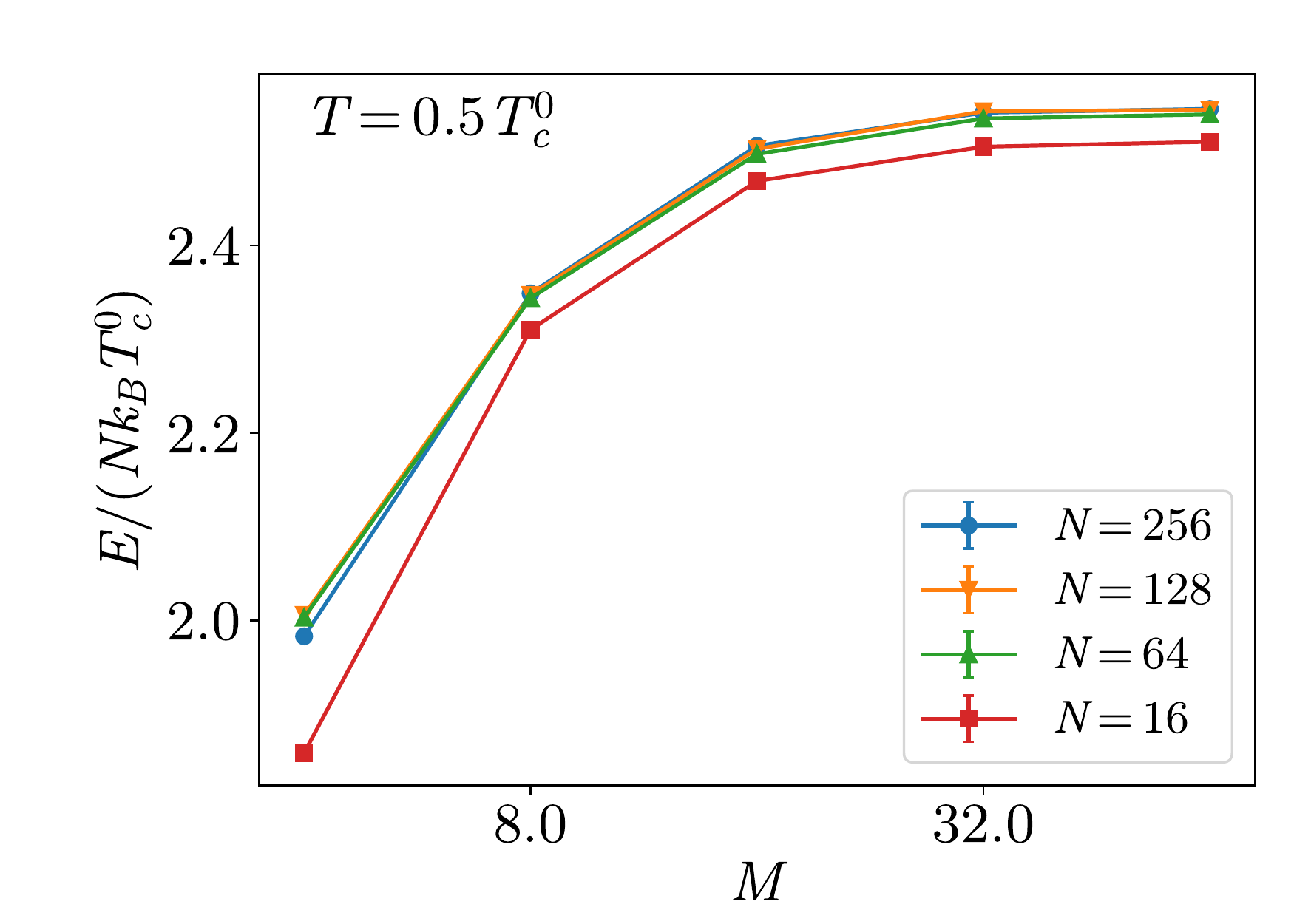}
	\caption{Internal energy $E$ for a system of hard-spheres bosons at the
		density $na^3=0.1$, shown as a function of the number of beads
		used in the
		simulation. The energy is computed using the virial estimator
		reported in
		appendix.
	}
	\label{fig:5}
\end{figure}

\subsection{Benchmarks}

\begin{figure}[t]
 \includegraphics[width=0.47\linewidth]{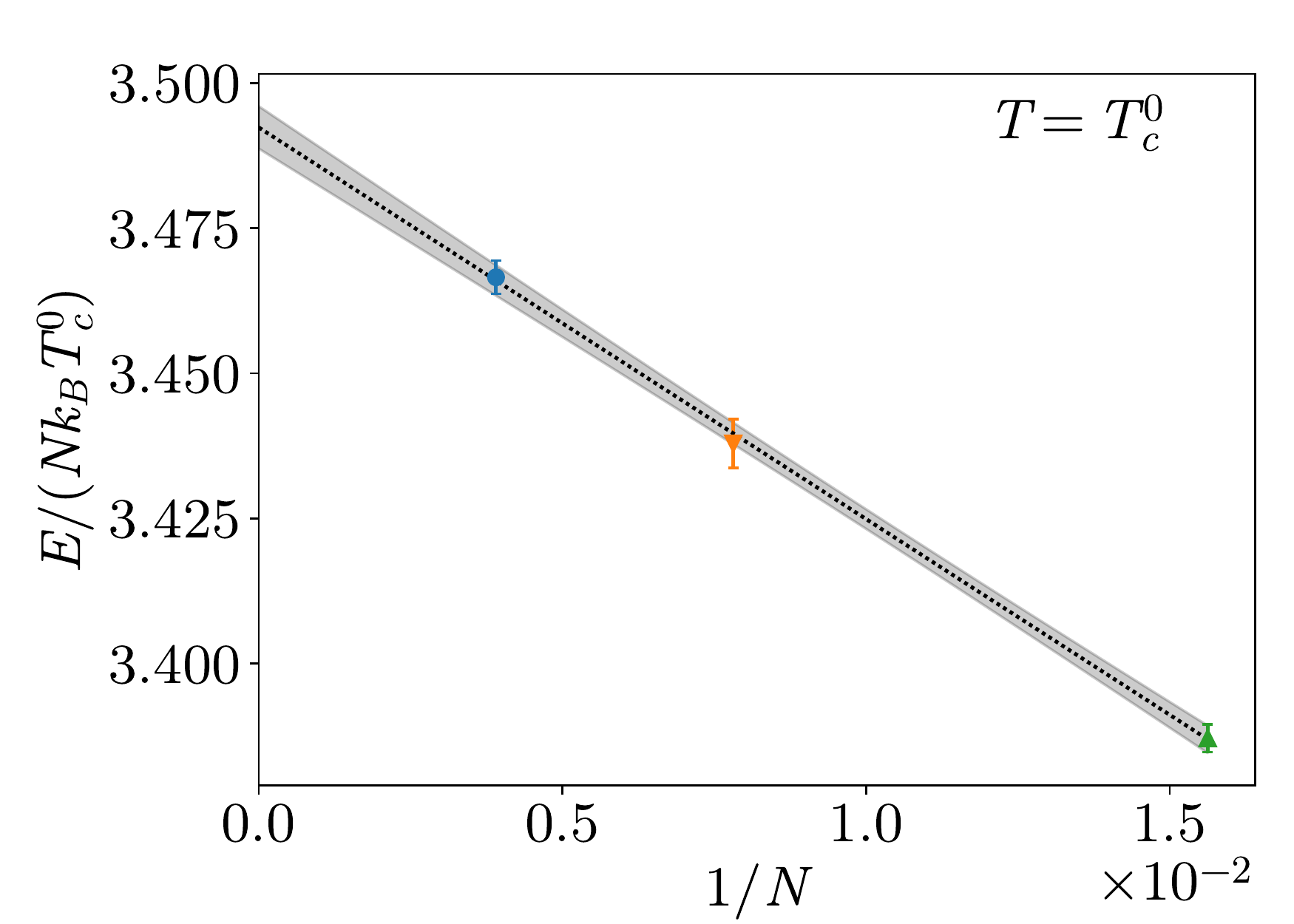}~~
 \includegraphics[width=0.47\linewidth]{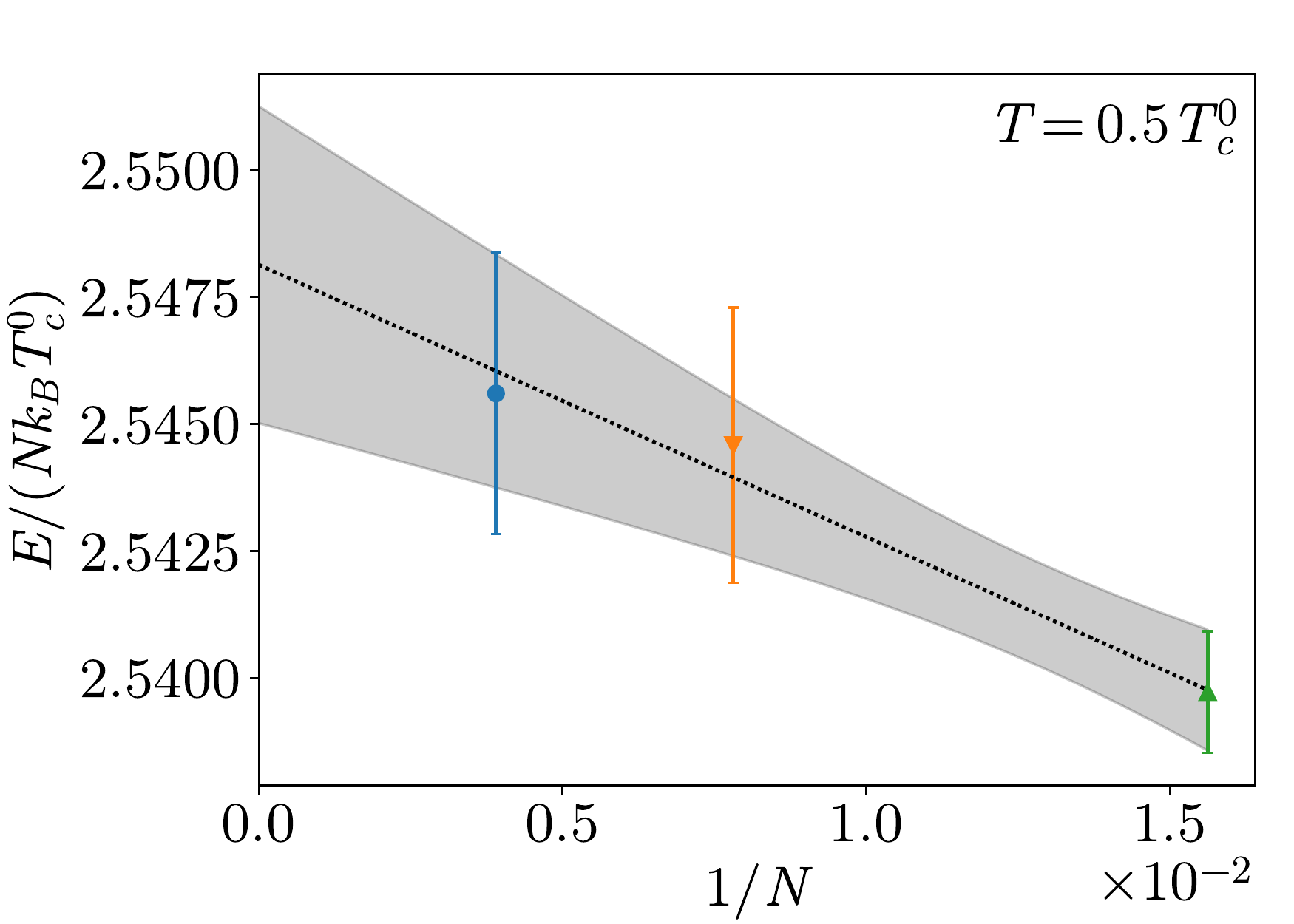}
 \caption{Extrapolation of the internal energy $E$ to the thermodynamic limit
  using the points at $M=64$ from fig.~\ref{fig:5}. We have omitted the system
  size $N=16$ because it is too small to be used in a linear extrapolation in
  $1/N$. Symbols follow the legend in fig.~\ref{fig:5}.}
 \label{fig:6}
\end{figure}

The presence of the hard-sphere potential, approximated by the product of
Cao--Berne density matrices brings a dependence on the number
of beads that is detectable when the gas parameter is large enough. In
fig.~\ref{fig:5} we show such dependence in a system at density $na^3=0.1$ for
two values of the temperature, $T=T_c^0$ and $T=0.5 T_c^0$. As one can
see, the value of the energy saturates at large number of beads, regardless of
the system size controlled by the total number of particles $N$. This can be
interpreted recalling that the Cao--Berne density matrix computes exactly
the scattering at high energy, requiring the thermal wavelength
associated with the imaginary time step $\dtau$ to be small compared to the
interaction range $a$, i.e.
\begin{equation}
 \sqrt{\frac{4\pi \lambda \beta}{M}} \ll a \,.
\end{equation}
On the contrary, in the limit of dilute systems the Cao--Berne
approximation becomes exact and even a small number of beads (even $8$ or
$16$) is sufficient to get precise results~\cite{PhysRevA.105.013325}.
Finally, in fig.~\ref{fig:6} we show how the points at $M=64$ from
fig.~\ref{fig:5} extrapolate to the thermodynamic limit for the two
temperatures. In the left and right panels of fig.~\ref{fig:6} the dotted lines
represent the linear fit to the data, with the gray bands covering the
one-sigma region. As expected, the system at the critical temperature $T_c^0$
suffers from much larger finite size effects, compared with the system at
temperature $T=0.5 T_c^0$.

\section{Conclusions}

We described an algorithm to perform unbiased PIMC simulations for Bose systems
in the canonical ensemble with periodic boundary conditions.
The formalism for path-integrals suitable for periodic systems has been
developed, and the technical details required for a correct implementation of
the PIMC algorithm have been discussed in a pedagogical manner. The formalism
can be similarly applied to the simulation of systems in the
grand-canonical ensemble.
Benchmark results have been presented for non-interacting Bose gases, for
which the energy can be exactly computed for any particle number.
Many-body systems with hard-sphere interactions have been addressed as well,
and the convergence to the continuous imaginary-time limit and to the infinite
system-size limit have been analyzed.
The PIMC algorithm we presented provides unbiased results for non-interacting
Bose gases with periodic boundary conditions for any number of imaginary-time
slices, meaning that, e.g., non-interacting particles can be exactly
simulated even	setting $M=1$.
Furthermore, all Monte Carlo updates can be performed without limitations on
the length of the paths that are involved in the update, even in regimes
where the thermal wavelength is comparable to the size of the fundamental
cell.
This is in contrast to previous implementations of the worm
algorithm~\cite{boninsegni2006worma}, for which, even for non-interacting
systems, unbiased results are obtained only for a sufficiently large number of
imaginary-time slices. Furthermore, those implementations become ambiguous when
the thermal wavelength starts to be comparable to the size of the fundamental
cell, possibly leading to biased results unless the length of the paths that
are involved in some updates is constrained.
For interacting systems, also the PIMC algorithm presented here requires a
sufficient number of imaginary-time slices, so that the high-temperature
approximation for the density matrix (in our case, the pair product
approximation) becomes essentially exact. However, there is no constraint on
the length of the  paths  involved in any updates.
It is worth emphasizing that the original implementation of the worm algorithm
provides unbiased results, even without constraints in the Monte Carlo
updates, if the size of the fundamental periodic cell is much larger than the
thermal wavelength. While such a system size is feasible for high and for
moderately low temperatures, it becomes impractical close to the zero
temperature limit.
When the thermal wavelength starts to be comparable to the size of the
fundamental periodic cell, stringent constraints in some Monte Carlo updates
have to be imposed in order to avoid ambiguities due to the choice of periodic
images. These constraints strongly reduce the acceptance rates, leading to
excessively long autocorrelation times and, therefore, to inefficient
simulations.
For these reasons, we expect the algorithm presented here to allow extending
the scope of the PIMC simulations, providing a better access to the intriguing
quantum phenomena occurring in the low temperature limit.
Furthermore, the possibility to perform exact benchmarks for small system sizes
will help novel practitioners in correctly implementing unbiased PIMC codes.

\begin{acknowledgments}
 This work was supported by the Italian Ministry of University and Research
 under the PRIN2017 project CEnTraL 20172H2SC4. S.P. acknowledges PRACE for
 awarding access to the Fenix Infrastructure resources at Cineca, which are
 partially funded from the European Union’s Horizon 2020 research and
 innovation
 programme through the ICEI project under the grant agreement No. 800858.
 S. P. also acknowledges the CINECA award under the ISCRA initiative, for the
 availability of high performance computing resources and support.
\end{acknowledgments}

\appendix

\section*{APPENDIX}

In this appendix we report the expressions needed to compute the internal
energy $E$ and pressure $P$ in a PIMC simulation, presenting two Monte Carlo
estimators for each of them.

\subsection{Energy}
The internal energy $E$ can be computed in the diagonal sector as
\begin{equation}
 E = -\frac{1}{Z_N}\frac{dZ_N}{d\beta} \,.
 \label{eq:E_def}
\end{equation}
Using the PIMC representation of $Z_N$ in eq.~\eqref{eq:Z_def2} and working out
the derivative, we obtain the thermodynamic estimator
\begin{equation}
 \frac{E_{\mathrm{th}}}{N} = \left< \frac{D}{2\dtau}
 -\frac{1}{4\lambda\dtau^2 N M} \sum_{j=0}^{M-1} \left(\Rv_j-\Rv_{j+1}\right)^2
 + \frac{1}{N M} \sum_{j=0}^{M-1} \frac{\partial
  U\left(\Rv_j,\Rv_{j+1}\right)}{\partial \dtau}
 \right>,
\end{equation}
where the average is taken on the configurations sampled in the $Z$-sectors. In
addition, the above quantity can be calculated in PIMC simulations using the so
called virial estimator~\cite{ceperley1995path}, which usually suffers from
smaller
statistical fluctuations. Different expressions can be derived, we report here
the version we have implemented in our code
\begin{multline}
 \frac{E_{\mathrm{vir}}}{N}
 = \Bigg< \frac{D}{2\beta}
 + \frac{\left(\Rv_{M-1} - \Rv_{M} \right) \cdot
  \left( \Rv_{M} - \Rv_0 \right)}{4\lambda\dtau^2 N M}
 + \frac{1}{2\beta N} \sum_{j=1}^{M-1} \left(\Rv_j- \Rv_0\right) \cdot
 \frac{\partial}{\partial \Rv_j}
 \left[ U\left(\Rv_{j-1},\Rv_{j}\right) + U\left(\Rv_{j},\Rv_{j+1}\right)
  \right] \\
 + \frac{1}{N M} \sum_{j=0}^{M-1} \frac{\partial
  U\left(\Rv_j,\Rv_{j+1}\right)}{\partial \tau} \Bigg> .
\end{multline}

\subsection{Pressure}

Although we have not presented any results for the pressure, we report for
completeness the expression for its two estimators. The pressure is defined as
\begin{equation}
 P = \frac{1}{\beta Z_N}\frac{dZ_N}{dV} \;.
 \label{eq:P_def}
\end{equation}
To calculate the derivative of the partition function with respect to the volume
$V$, we use the identity $V d\Rv_j/dV = \Rv_j / D$ and apply the chain rule for
every $j$. The thermodynamic estimator reads
\begin{equation}
 \frac{P_{\mathrm{th}}}{n} =
 \Bigg<
 \frac{1}{\dtau}
 - \frac{1}{2 \lambda \dtau^2 N M D} \sum_{j=0}^{M-1}
 \left(\Rv_j-\Rv_{j+1}\right)^2
 - \frac{V}{\beta N} \sum_{j=0}^{M-1}
 \frac{\partial U(\Rv_j, \Rv_{j+1})}{\partial V}
 \Bigg> \,.
\end{equation}
Notice that we have not used the chain rule on the last term because, depending on the
interatomic potential and on the adopted approximation for the density matrix, it is 
sometimes easier to directly compute
the derivative of the interaction term (e.g.~the expression in eq.~\eqref{eq:U_CaoBerne})
with respect to the volume, obtaining an expression without ambiguities due to periodic boundary
conditions. The virial estimator can instead be computed as
\begin{multline}
 \frac{P_{\mathrm{vir}}}{n} =
 \Bigg<
 \frac{1}{\beta}
 + \frac{\left( \Rv_{M-1}-\Rv_{M} \right) \cdot
  \left( \Rv_{M}-\Rv_0 \right)}{2\lambda \dtau^2 N M  D}
 + \frac{1}{\beta N D} \sum_{j=1}^{M-1} \left(\Rv_j- \Rv_0\right) \cdot
 \frac{\partial}{\partial \Rv_j}
 \left[ U\left(\Rv_{j-1},\Rv_{j}\right) + U\left(\Rv_{j},\Rv_{j+1}\right)
  \right]\\
 - \frac{V}{\beta N} \sum_{j=0}^{M-1}
 \frac{\partial U(\Rv_j, \Rv_{j+1})}{\partial V}
 \Bigg>.
\end{multline}

\bibliography{refs}

\end{document}